\def\year{2016}\relax

\documentclass[letterpaper]{article}
\usepackage{aaai16}
\usepackage{times}
\usepackage{helvet}
\usepackage{courier}
\usepackage{graphicx}
\usepackage{url}
\usepackage[
  bookmarks=false,
  pdfpagelabels=false,
  hyperfootnotes=false,
  hyperindex=false,
  pageanchor=false,
]{hyperref}
\usepackage{makecell}
\usepackage{balance}
\makeatletter
\begin{document}

% ==================================================== %
%               Title and Abstract                     %
% ==================================================== %

\title{A Data-driven Study of View Duration on YouTube \\ \scalebox{.7}{[Please cite the ICWSM'16 version of this paper]}}
\author{Minsu Park$^1$, Mor Naaman$^1$, Jonah Berger$^2$\\ $^1$Jacobs Institute, Cornell Tech\\ $^2$The Wharton School, University of Pennsylvania\\ \{minsu, mor\}@jacobs.cornell.edu\\ jberger@wharton.upenn.edu}
\maketitle

\begin{abstract}
Video watching had emerged as one of the most frequent media activities on the Internet. Yet, little is known about how users watch online video. Using two distinct YouTube datasets, a set of random YouTube videos crawled from the Web and a set of videos watched by participants tracked by a Chrome extension, we examine whether and how indicators of collective preferences and reactions are associated with view duration of videos. We show that video view duration is positively associated with the video's view count, the number of likes per view, and the negative sentiment in the comments. These metrics and reactions have a significant predictive power over the duration the video is watched by individuals. Our findings provide a more precise understandings of user engagement with video content in social media beyond view count.
\end{abstract}

% ==================================================== %
%               Introduction                           %
% ==================================================== %

\section{Introduction}
\noindent Video watching is perhaps the most popular web-based activity, through video hosting and sharing services such as YouTube, Facebook, Netflix, Vimeo, and others~\cite{cisco2015traffic}. As of 2015, YouTube alone has more than 1 billion viewers every day, watching hundreds of millions of hours of content~\cite{youTube2015stats}. It is forecasted that video will represent 80 percent of all traffic by 2019~\cite{cisco2015traffic}. Yet, little is known about how users engage with and watch online video. We use two distinct datasets from YouTube to investigate how users' engagement in watching a video (i.e., view duration) is associated with other video metrics such as the number of views, likes, comments, and the sentiment of comments.

A number of research efforts have investigated view count as a key indicator of popularity or quality of video---particularly looking at its relationships with other popularity or preference metrics (e.g., the number of likes and comments). For example, the number of comments/favorites/ratings and average rating are significant predictors of video view counts on YouTube~\cite{chatzopoulou2010first}; the sequence of comments and its structure are strongly associated with view counts~\cite{de2009makes}; and view counts can be predicted through socially shared viewing behaviors around the content such as how many times a video was rewound or fast-forwarded as well as the duration of the session in a tool that allows people watch videos together in sync and real time~\cite{shamma2011viral}.

Although views, likes, comments, and other such measures can be considered as indicators of general popularity and preferences, there has been growing interest in using deeper post-click user engagement (e.g., how long a user watched a video) to estimate more accurate relevance and interest and to improve ranking and recommendation~\cite{yi2014beyond}. For example, YouTube has started to use `dwell time' (the length of time that a user spends on a video, e.g., video watching session length) instead of click events to better measure the engagement with video content~\cite{youtube2015dwell}. Beyond video, Facebook is using dwell time on external links to combat Clickbait---stories with arousing headlines that attract users to click and share more than usual, but are not consumed in depth~\cite{cramer2015effects,El-Arini2014clickbait}.

Understanding the relationships between watching behavior and other popularity and engagement metrics can help develop more comprehensive behavioral models of user engagement and preferences beyond view count. In this work, we explore the relationship between user engagement with video (i.e., the portion of the video watched by the user) and other activity metrics (e.g., the number of views/likes/dislikes/comments) that are often considered as indicators of popularity or collective preference~\cite{baym2013data} and may impact or at least align with user engagement and watch duration. In fact, our basis hypothesis is that:

\textit{H1: Video view duration is positively associated with other video popularity and quality measures.}

Moreover, we examine another factor which may be associated with video view duration: the sentiment of video comments. A number of recent studies in different disciplines provided potential links among emotionality of contents, comments, and user engagement. Comments are indicative of users' opinion of and attitude towards video content~\cite{yew2011knowing}. Interest, enjoyment, and desire to find out more about online news are enticed when strong sentiment and negative connotations are present in the content~\cite{arapakis2014user}. Further, emotionality of content and its intensity can be reflected on comments~\cite{alhabash2015comment}, or, alternatively, certain aspects of content may cause people to comment with strong sentiment. In each case, we expect that sentiment of comments are correlated with engagement in terms of view duration:

\textit{H2: Video view duration is positively associated with the sentiment (positive or negative) of comments.}

To examine how those indicators of collective preferences and reactions are associated with engagement, we devise a data-driven study, using two distinct YouTube datasets: a set of randomly selected YouTube videos crawled from the Web and a set of videos watched by study participants that installed a Chrome extension tracking their behavior on YouTube. The main contribution in this work, then, is showing the robust relationships between view count, the number of likes per view, negative sentiment in comments, and video view duration. In this, we show that observed popularity metrics indeed have a significant predictive power for engagement. These findings could inform web services that require more precise understandings of engagement with social media video content beyond view count.

% ==================================================== %
%               Method                                 %
% ==================================================== %

\section{Method}
\noindent To make our findings and analysis more robust, we used two distinct datasets.\footnote{Both datasets are available on
%\url{http://github.com/sTechLab/YouTubeDurationData}}
\href{http://github.com/sTechLab/YouTubeDurationData}{http://github.com/sTechLab/ YouTubeDurationData}}
For each of these datasets we computed an overlapping but not completely identical set of variables. In particular, the dependent variable representing view duration is different in both datasets, as we detail below. 

\subsection{Data Collection}
\subsubsection{Random Videos.} This dataset is a sample of 1,125 random videos from YouTube. We first obtained 100,000 randomly sampled YouTube videos through Random YouTube, a site that collects IDs of the 10 most recently uploaded videos every fifteen minutes.\footnote{
%\url{http://randomyoutube.net}}
\href{http://randomyoutube.net}{http://randomyoutube.net}}
We then retrieved the video properties for each of the videos through the YouTube API. To find the average watch duration for each video, we scraped the YouTube video page to get the aggregate watch duration available on the video page statistics tab using tools from~\cite{yu2015lifecyle}. This step resulted in 44,766 videos---other videos were either private videos, were deleted, or did not provide aggregate statistics. Finally, to ensure proper measurement of sentiment, we required videos in our sample to have at least five English comments, resulting in a final set of 1,125 videos.

For the \emph{Random Videos} dataset, our view duration dependent variable was computed using the video's \textit{average view duration}: the aggregate view duration of the video (as reported by YouTube) divided by the view count, both reported over the video's lifetime.

\subsubsection{Individual Logs.} The \emph{Individual Logs} dataset is a set of 1,814 videos watched over several weeks by a sample of 158 participants recruited from Task Rabbit.\footnote{
%\url{http://taskrabbit.com}}
\href{http://taskrabbit.com}{http://taskrabbit.com}}
To obtain this individual-level data, we developed and deployed a Chrome extension that collected how long each user spent each YouTube video page they visited, as well as other YouTube video information available at the time of the viewing. The extension recorded a view timestamp, video properties, and dwell time in each viewing session. The extension was installed and used by 189 participants who had a previous experience with YouTube, over a span of at least two weeks (our sample included 77 male and 112 female participants; there were no significant gender differences in usage of the extension). Using this extension, we collected a total of 17,599 video view sessions. We filtered videos which had missing information, were private, deleted, or did not have at least five English comments. We only kept videos where the participant's dwell time was not greater than the video duration to avoid unusually long dwell time due to potential errors (e.g., unexpected disconnections).\footnote{We examined different cutoffs on dwell time from 1.5 times to 3 times longer than the video duration with truncation at 1 and found no impact on our key findings.} The final dataset, as mentioned above, included 1,814 video view sessions.

For the \emph{Individual Logs} dataset, our view duration dependent variable was computed differently. In this case, we have used an individual, but approximate, view duration measurement. In particular, we used the user's dwell time for each video on the video's page, as was measured by the extension, as an approximation for the actual view time for the video by that user. 

To summarize, we used two distinct datasets and collected or computed variables in slightly different fashion for each. The \emph{Random Videos} dataset provides an aggregate understanding of average view duration that may capture life cycle dynamics of videos (e.g., older videos are potentially viewed differently over time) whereas the \emph{Individual Logs} dataset provides an understanding of individual's viewing behavior. Note that videos viewed by our participants tend to be more popular than the videos in the \emph{Random Videos} dataset. As we show below, while these two datasets resulted in a different distribution of videos as reflected in the video metadata variables, our results, associating these variables with the dependent variable of view duration, are fairly robust across datasets, adding confidence in the outcome.

%% Gilbert Histogram %%

\begin{table}[!t]
\centering
  \def\arraystretch{1.2} % 1 is the default
\resizebox{\columnwidth}{!}{%
\begin{tabular}{lllll}

\textbf{{\fontfamily{pag}\selectfont }} & \textbf{{\fontfamily{pag}\selectfont Random Videos}} & \textbf{{\fontfamily{pag}\selectfont Max}} & \textbf{{\fontfamily{pag}\selectfont Individual Logs}} &  \textbf{{\fontfamily{pag}\selectfont Max}}\\ \Xhline{2\arrayrulewidth}
{\fontfamily{pag}\selectfont View Duration Prop.} & \raisebox{-.2\height}{\includegraphics[width=3.0cm,height=0.8cm]{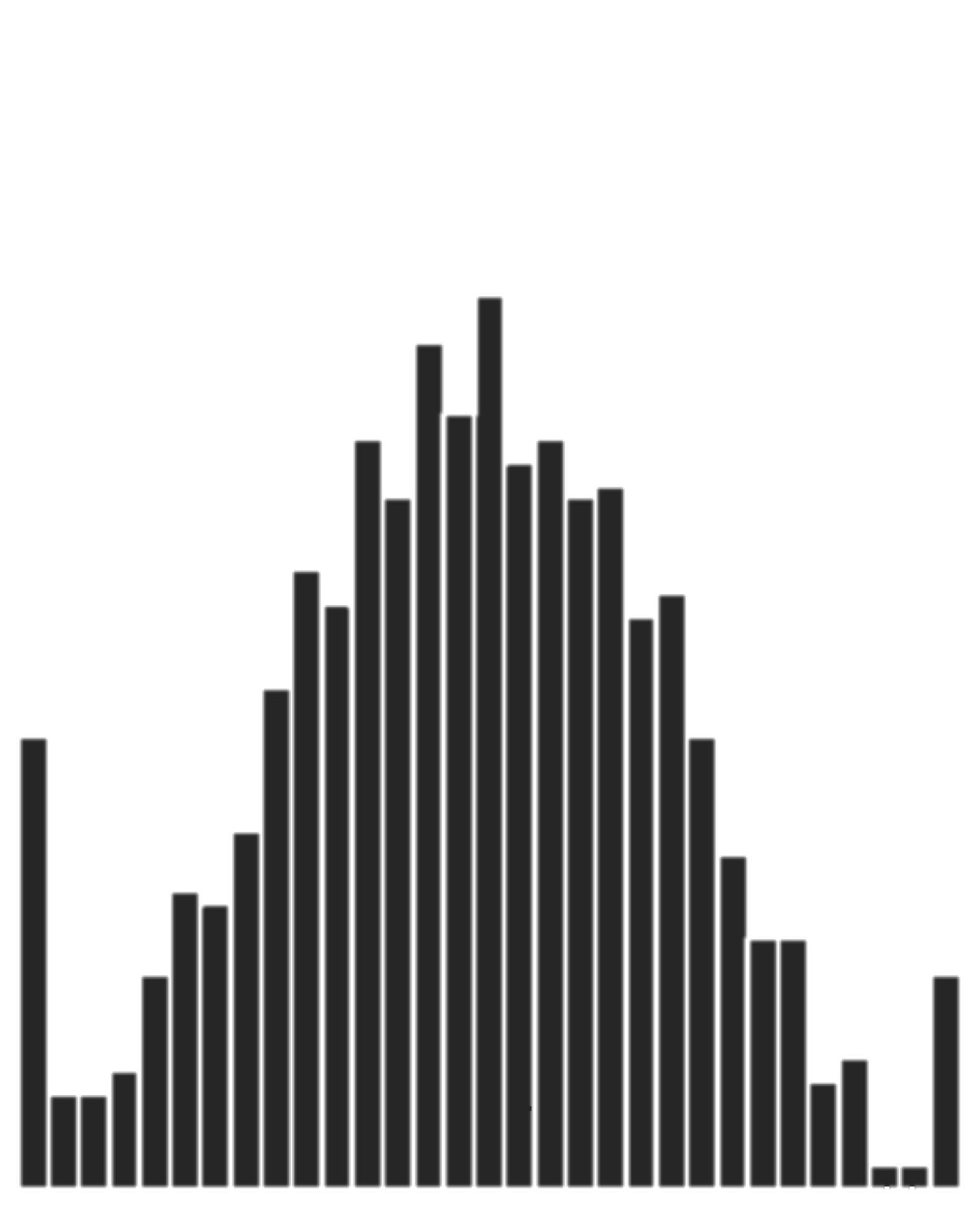}} & {\fontfamily{pag}\selectfont 1} & \raisebox{-.2\height}{\includegraphics[width=3.0cm,height=0.8cm]{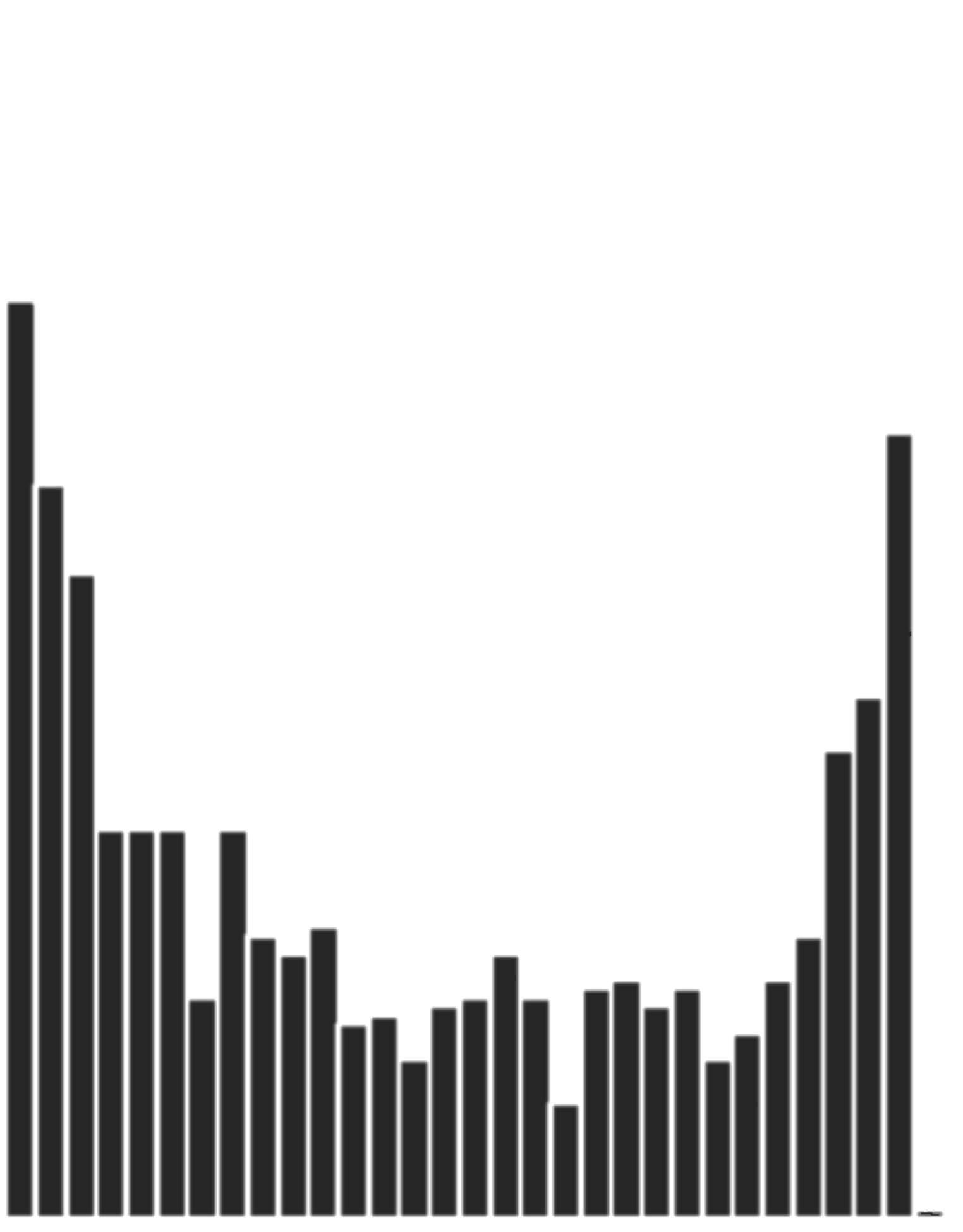}} & {\fontfamily{pag}\selectfont 1}\\
\\
\textbf{{\fontfamily{pag}\selectfont Video Properties}} & & \\ \Xhline{2\arrayrulewidth}
{\fontfamily{pag}\selectfont Duration (Sec)} &  \raisebox{-.2\height}{\includegraphics[width=3.0cm,height=0.8cm]{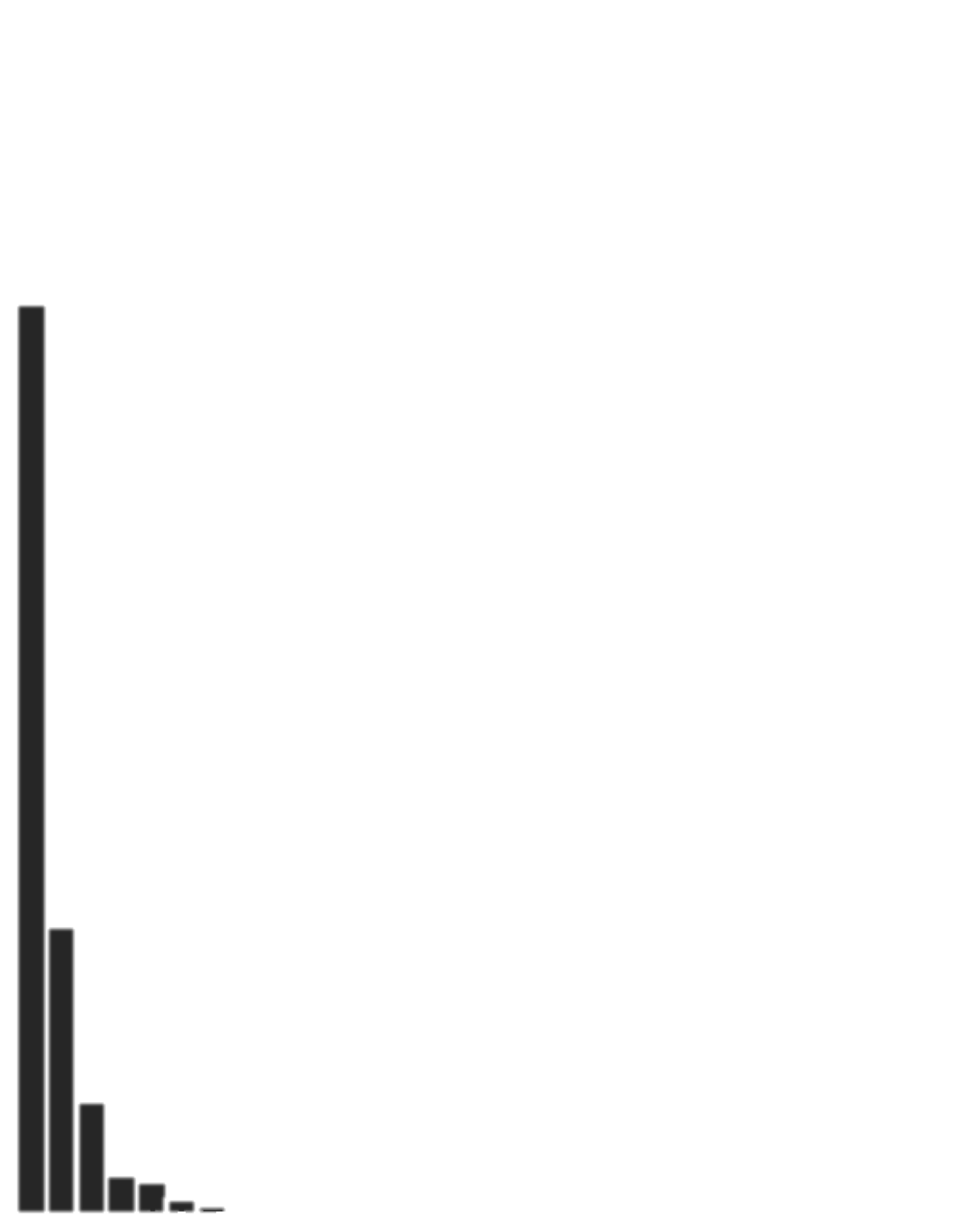}} & {\fontfamily{pag}\selectfont 15,229} & \raisebox{-.2\height}{\includegraphics[width=3.0cm,height=0.8cm]{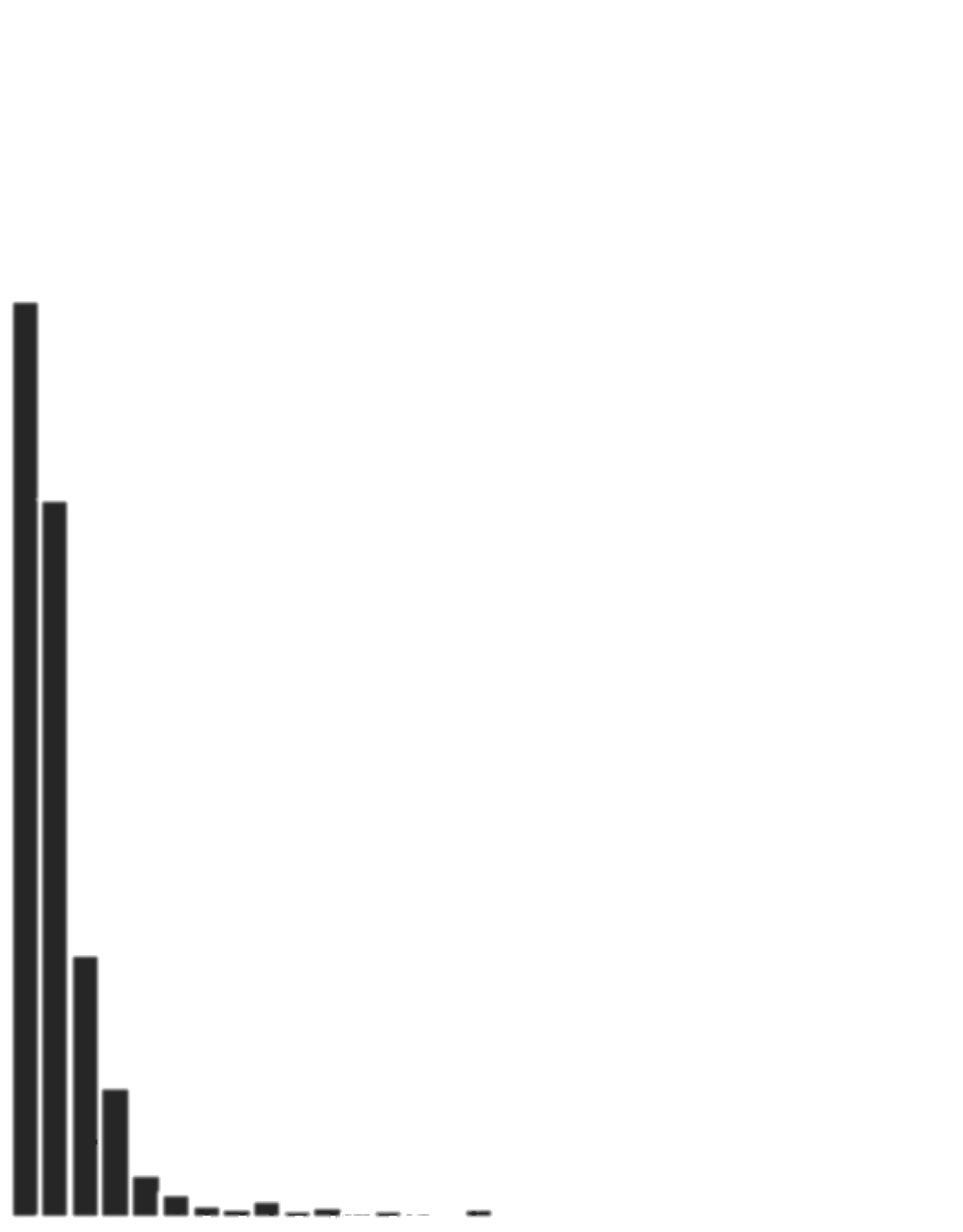}} & {\fontfamily{pag}\selectfont 7,267}\\
{\fontfamily{pag}\selectfont Elapsed Time (Days)} &  \raisebox{-.2\height}{\includegraphics[width=3.0cm,height=0.8cm]{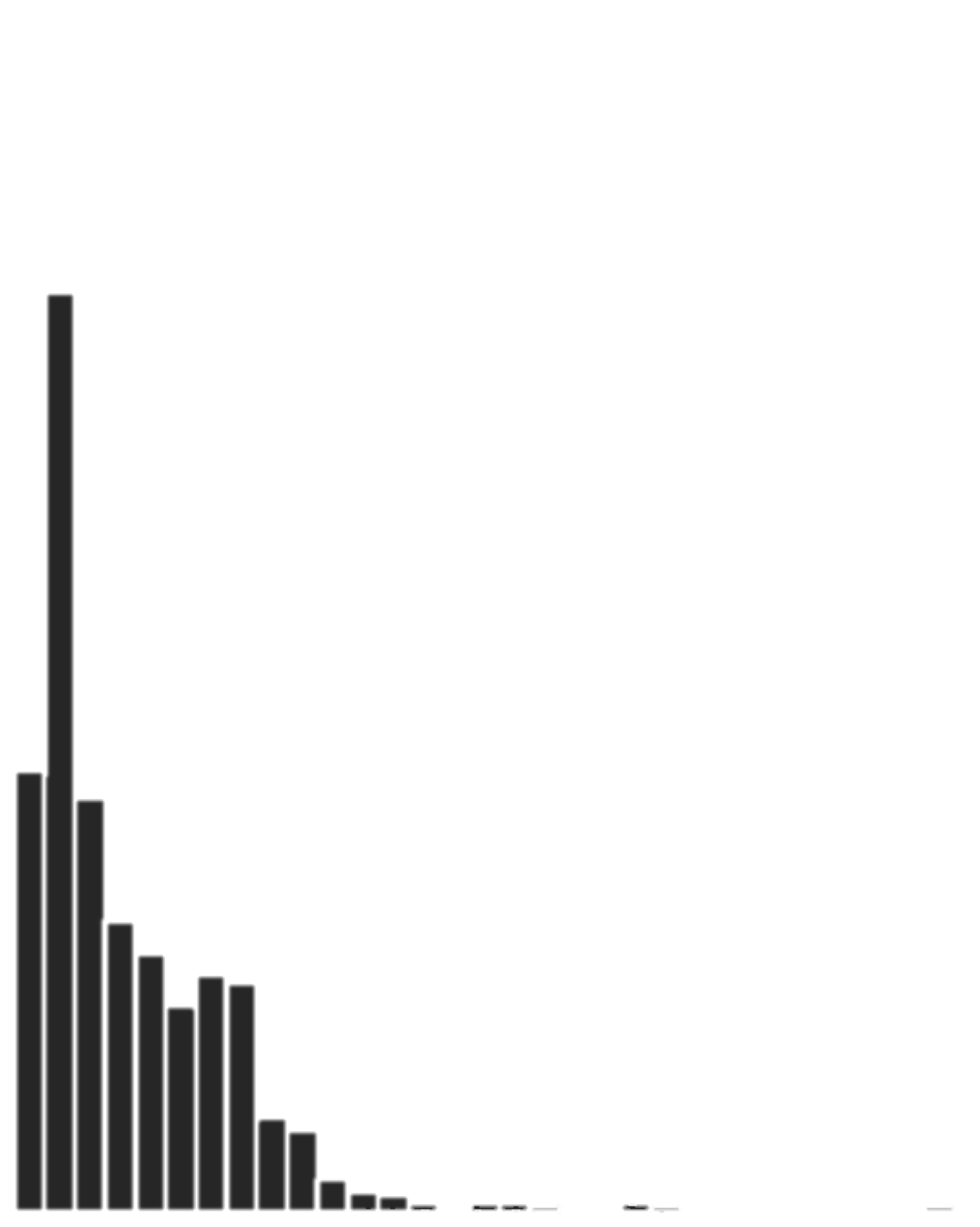}} & {\fontfamily{pag}\selectfont 3,059} & \raisebox{-.2\height}{\includegraphics[width=3.0cm,height=0.8cm]{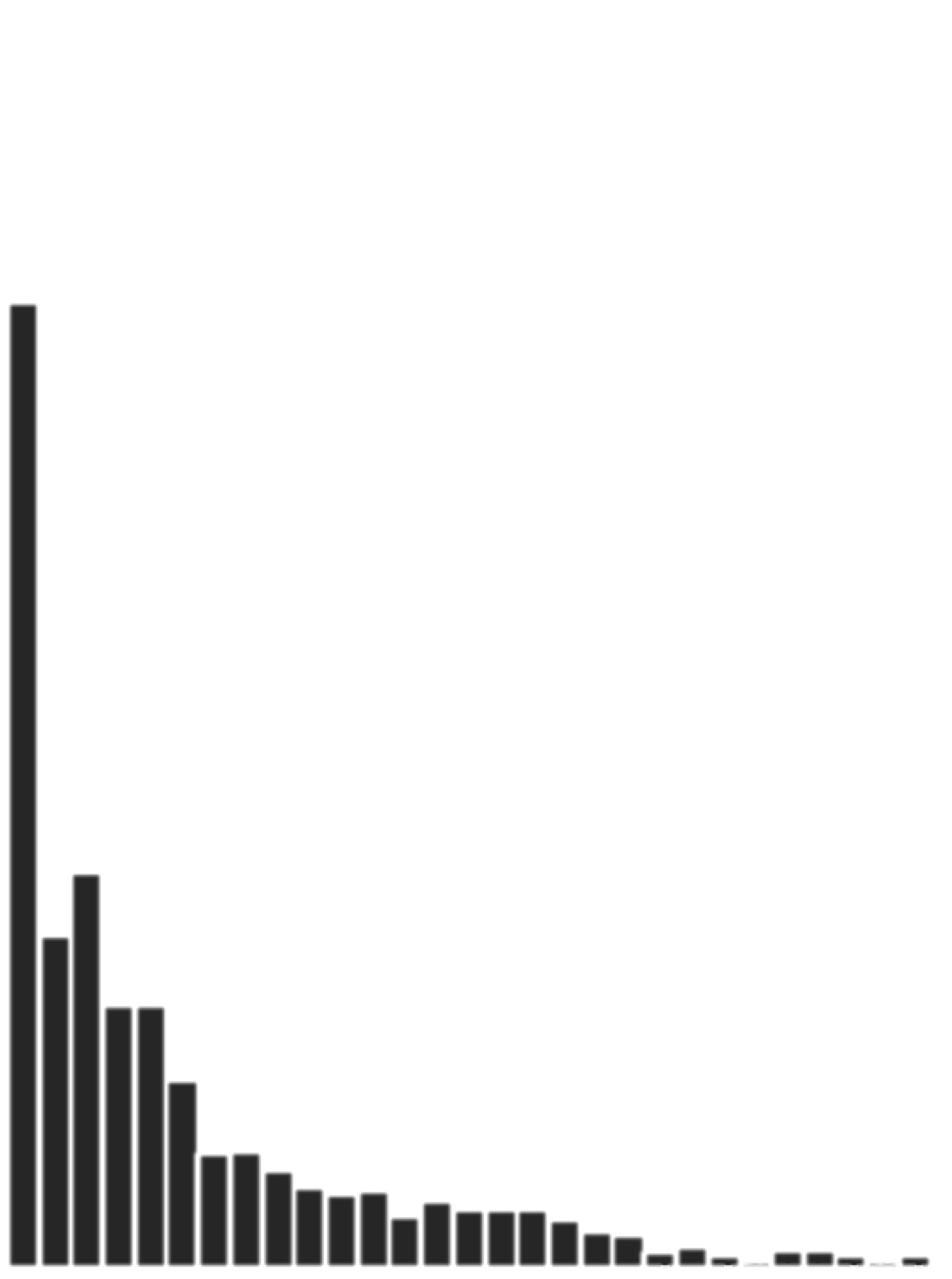}} & {\fontfamily{pag}\selectfont 3,140}\\ 
\\
\textbf{{\fontfamily{pag}\selectfont Metrics \& Sentiments}} & & \\ \Xhline{2\arrayrulewidth}
{\fontfamily{pag}\selectfont View Count} &  \raisebox{-.2\height}{\includegraphics[width=3.0cm,height=0.8cm]{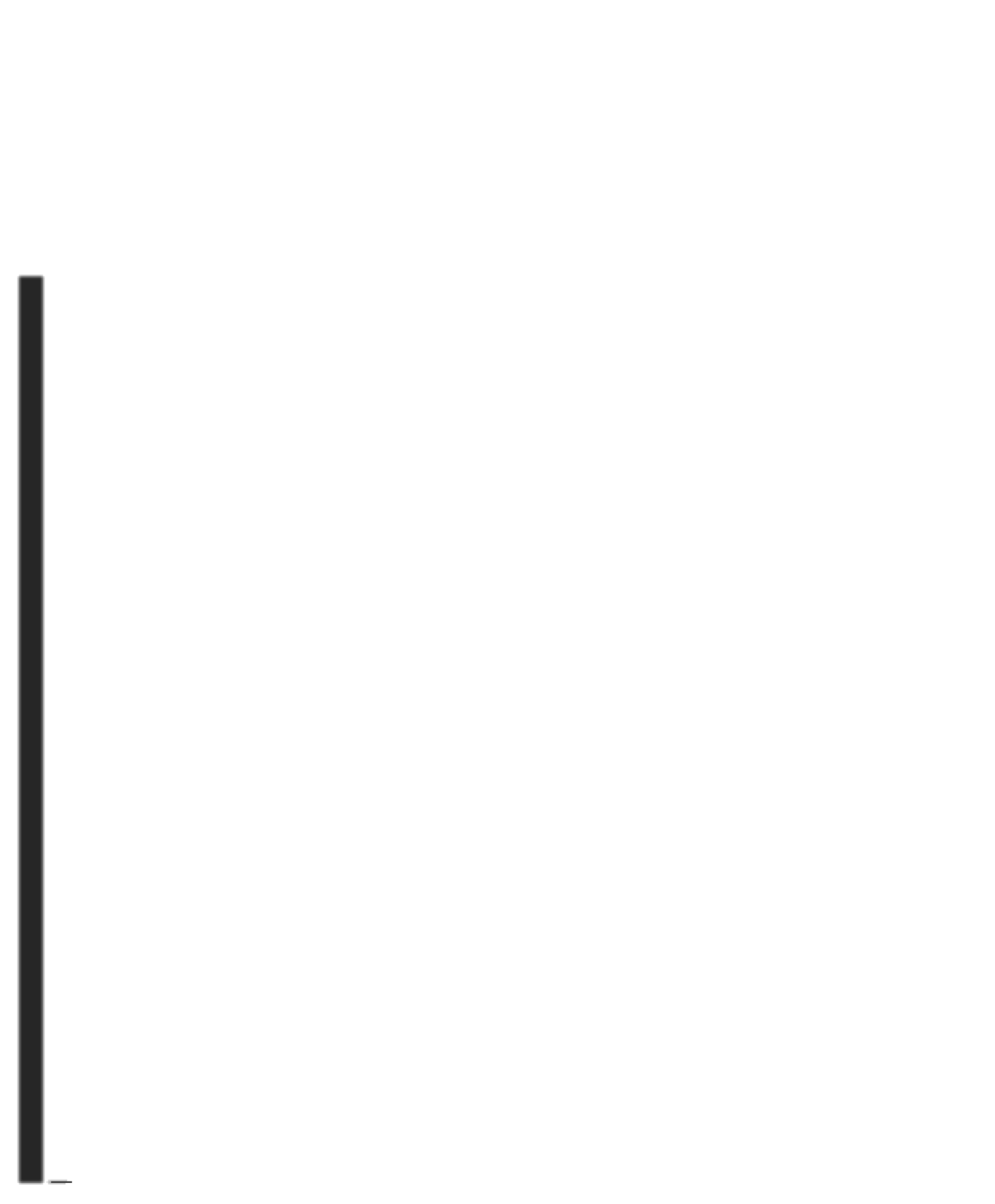}} & {\fontfamily{pag}\selectfont 56M} & \raisebox{-.2\height}{\includegraphics[width=3.0cm,height=0.8cm]{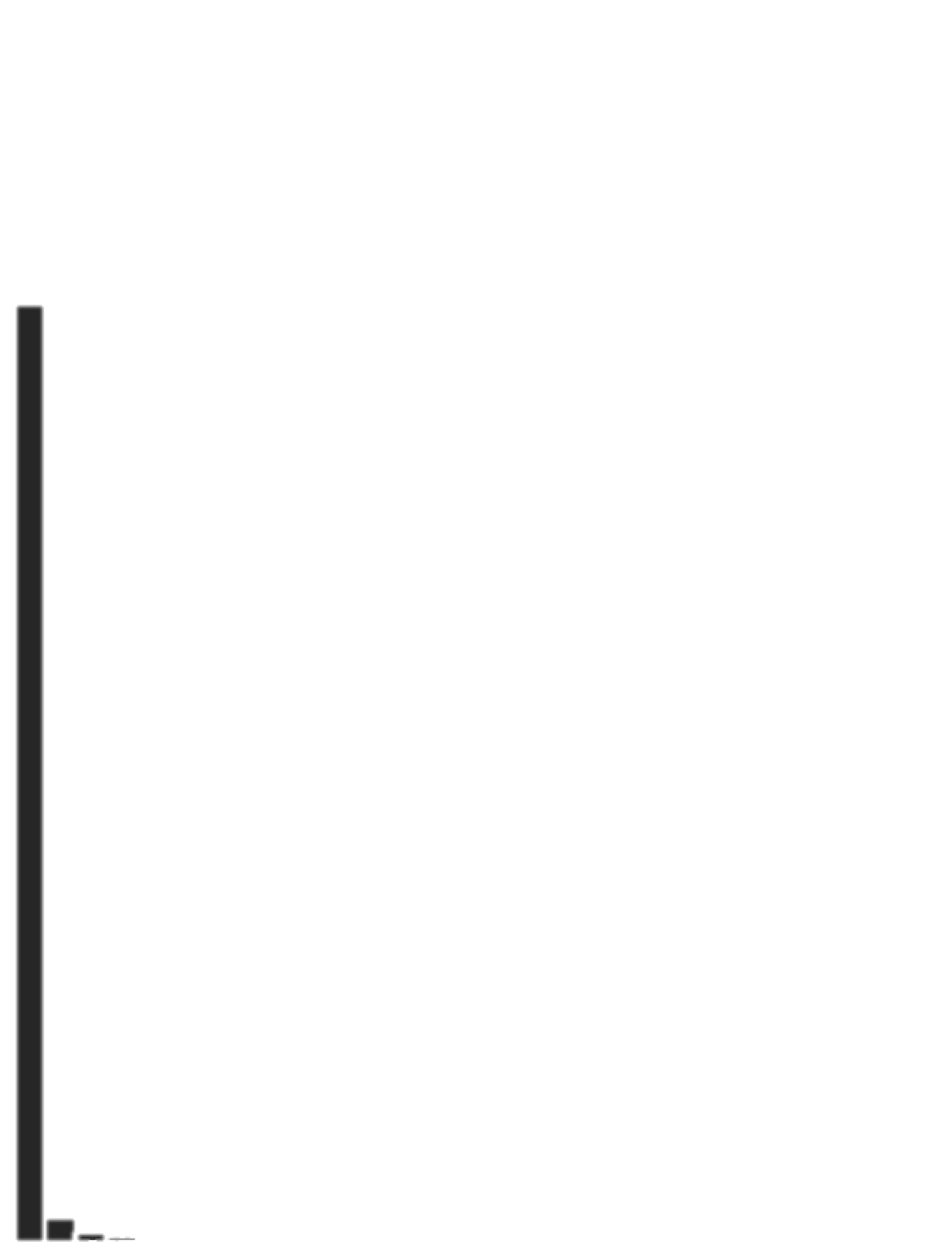}} & {\fontfamily{pag}\selectfont 678M}\\
{\fontfamily{pag}\selectfont Comments per View} &  \raisebox{-.2\height}{\includegraphics[width=3.0cm,height=0.8cm]{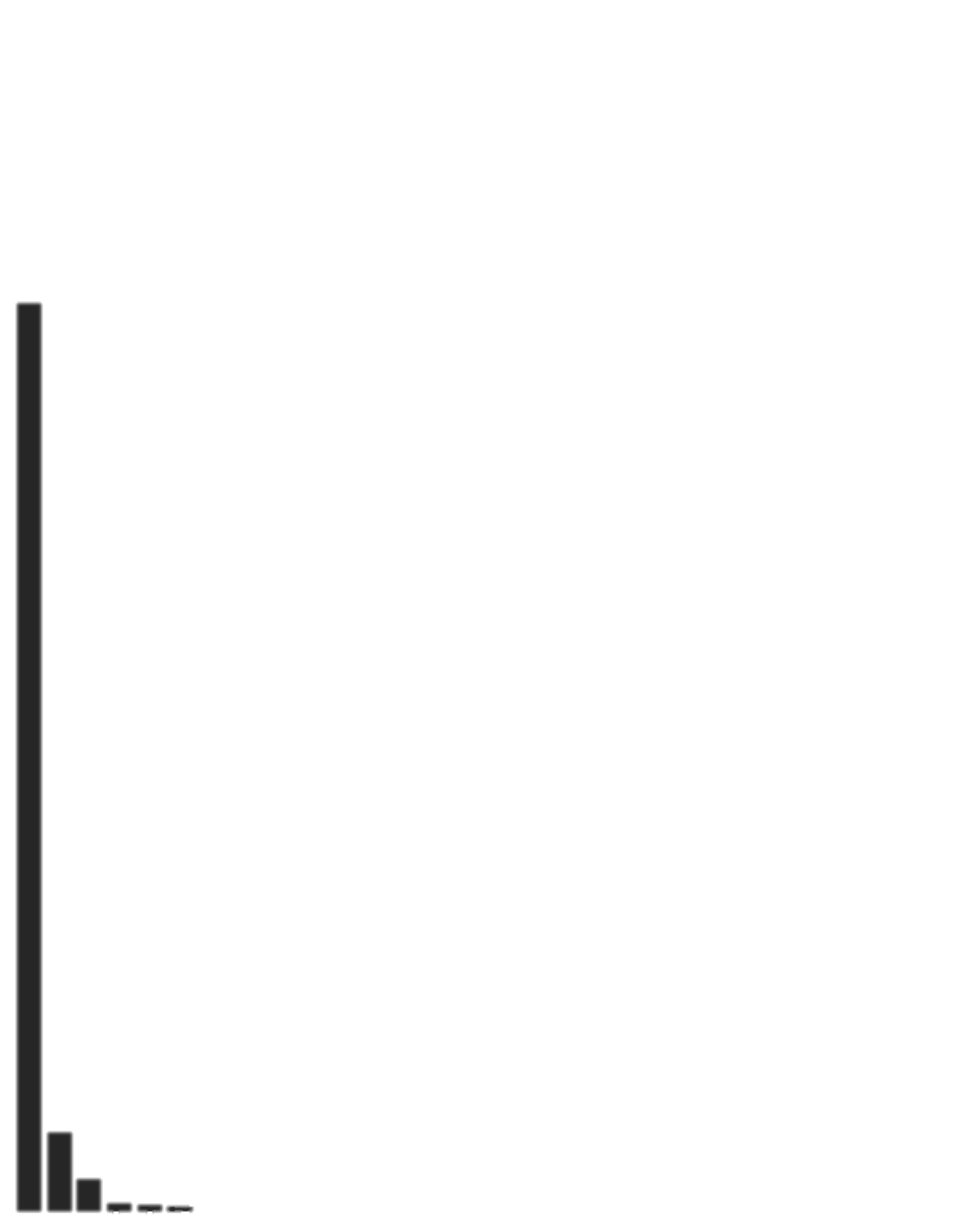}} & {\fontfamily{pag}\selectfont 1.33} & \raisebox{-.2\height}{\includegraphics[width=3.0cm,height=0.8cm]{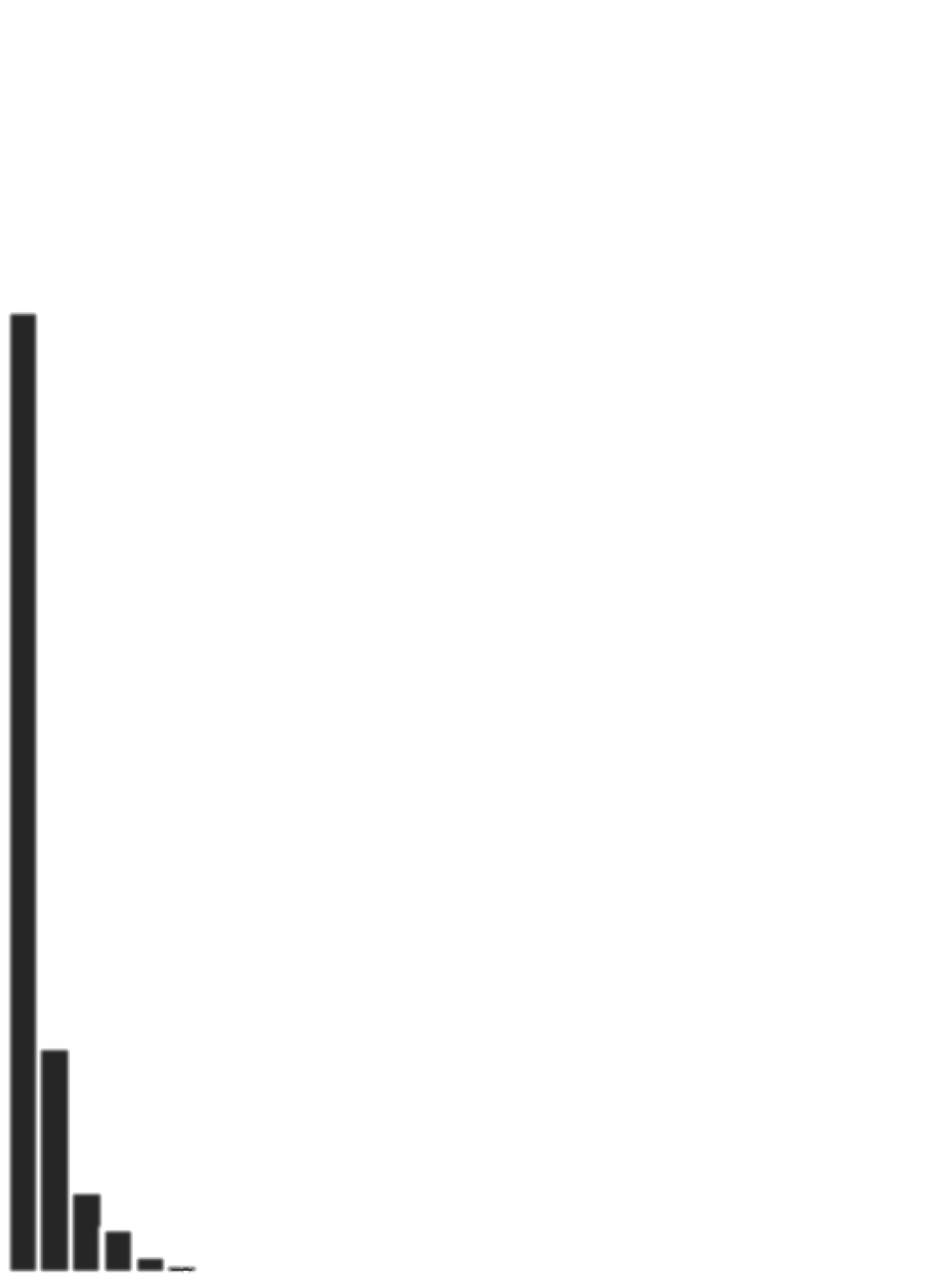}} & {\fontfamily{pag}\selectfont 0.075}\\
{\fontfamily{pag}\selectfont Discussion Density} &  \raisebox{-.2\height}{\includegraphics[width=3.0cm,height=0.8cm]{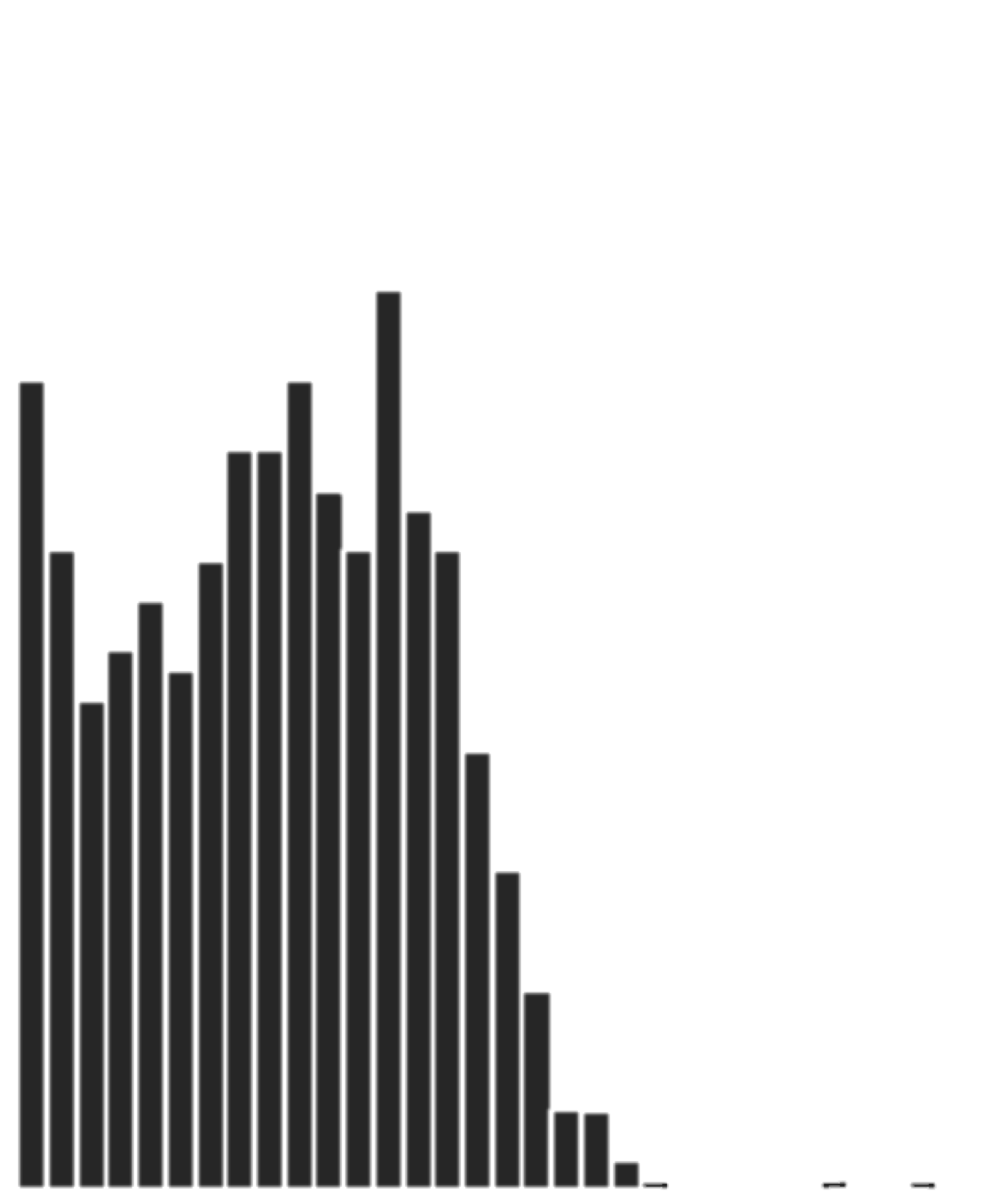}} & {\fontfamily{pag}\selectfont 1.11} & \raisebox{-.2\height}{\includegraphics[width=3.0cm,height=0.8cm]{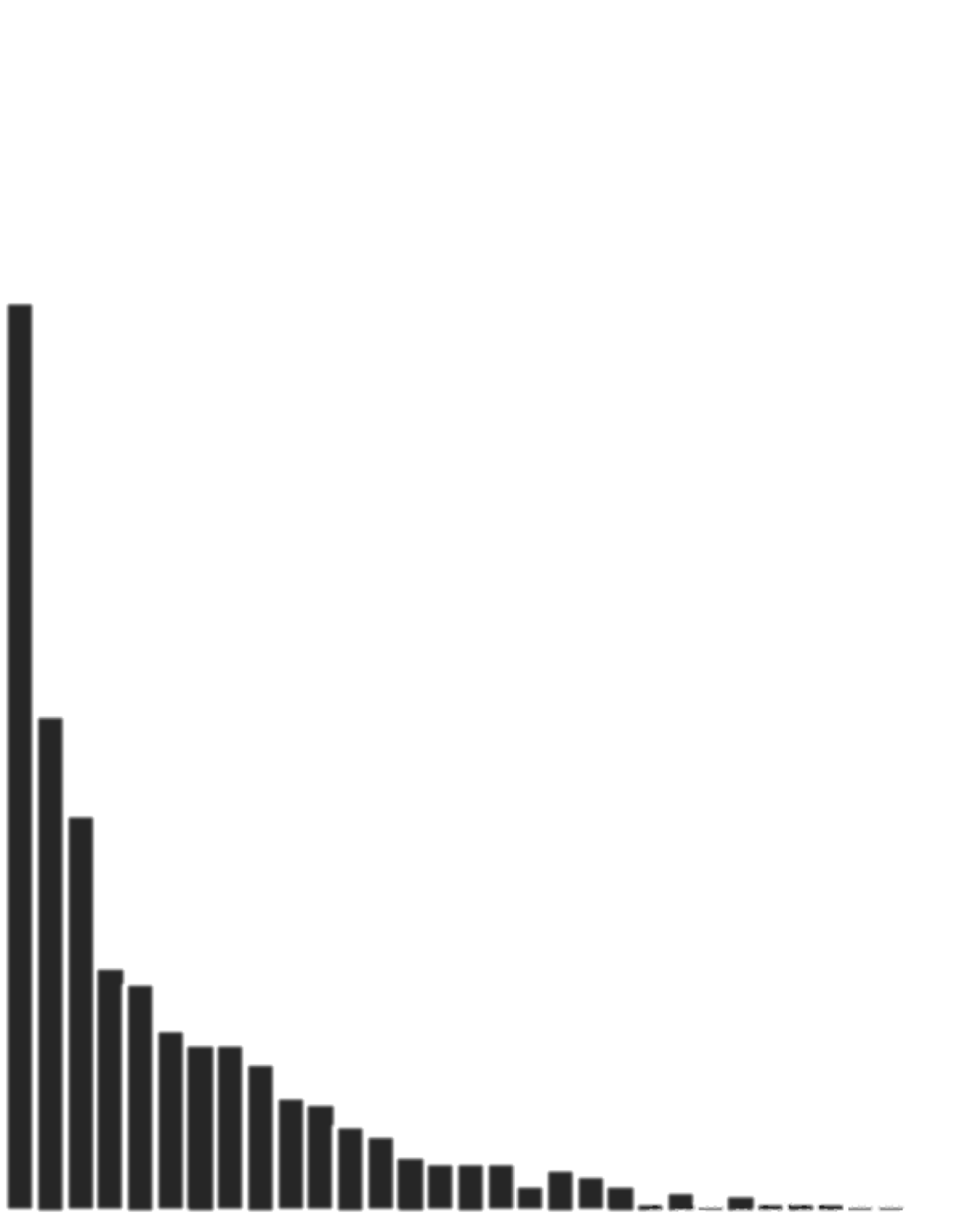}} & {\fontfamily{pag}\selectfont 0.71}\\
{\fontfamily{pag}\selectfont Positive Emotion} &  \raisebox{-.2\height}{\includegraphics[width=3.0cm,height=0.8cm]{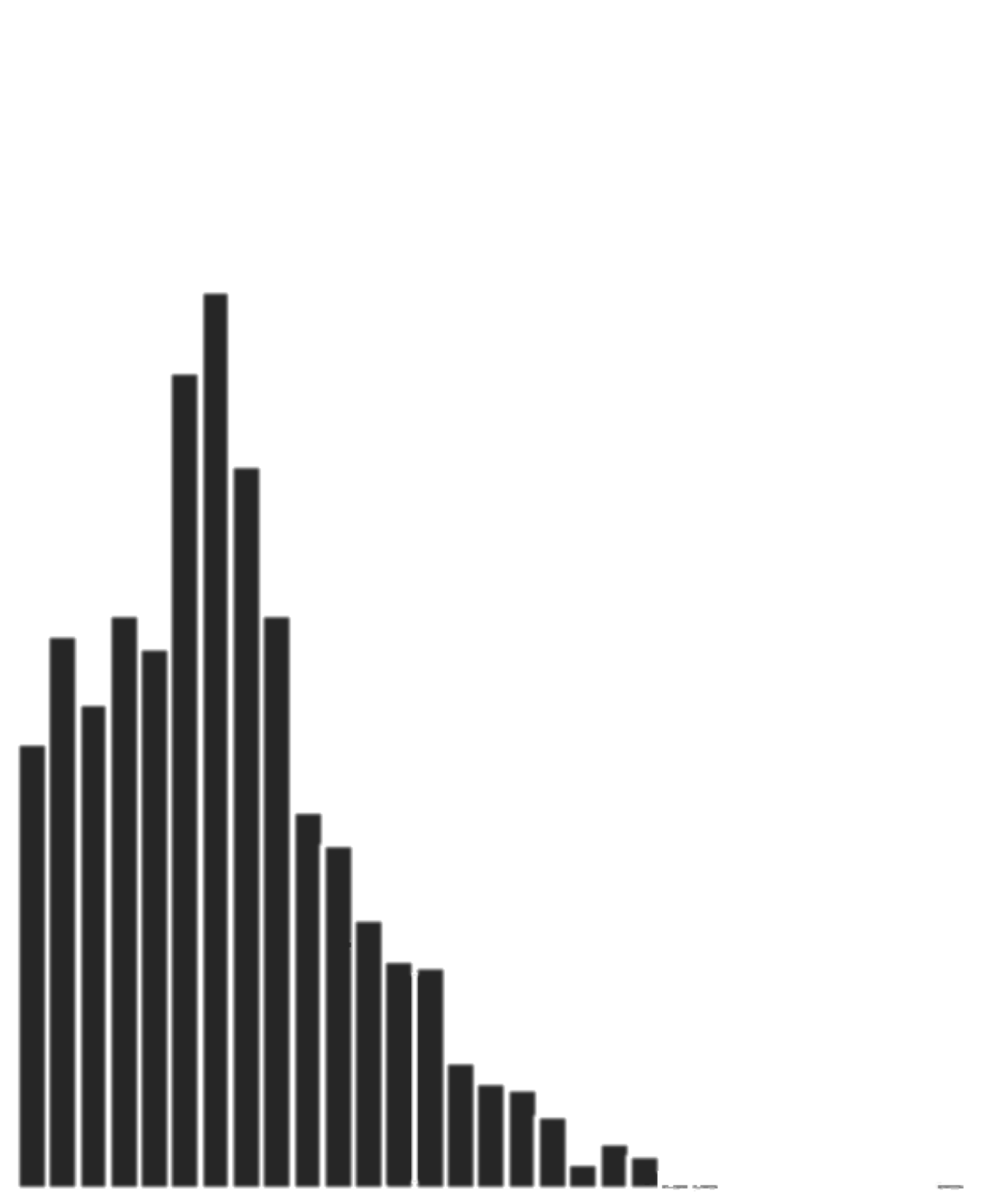}} & {\fontfamily{pag}\selectfont 0.84} & \raisebox{-.2\height}{\includegraphics[width=3.0cm,height=0.8cm]{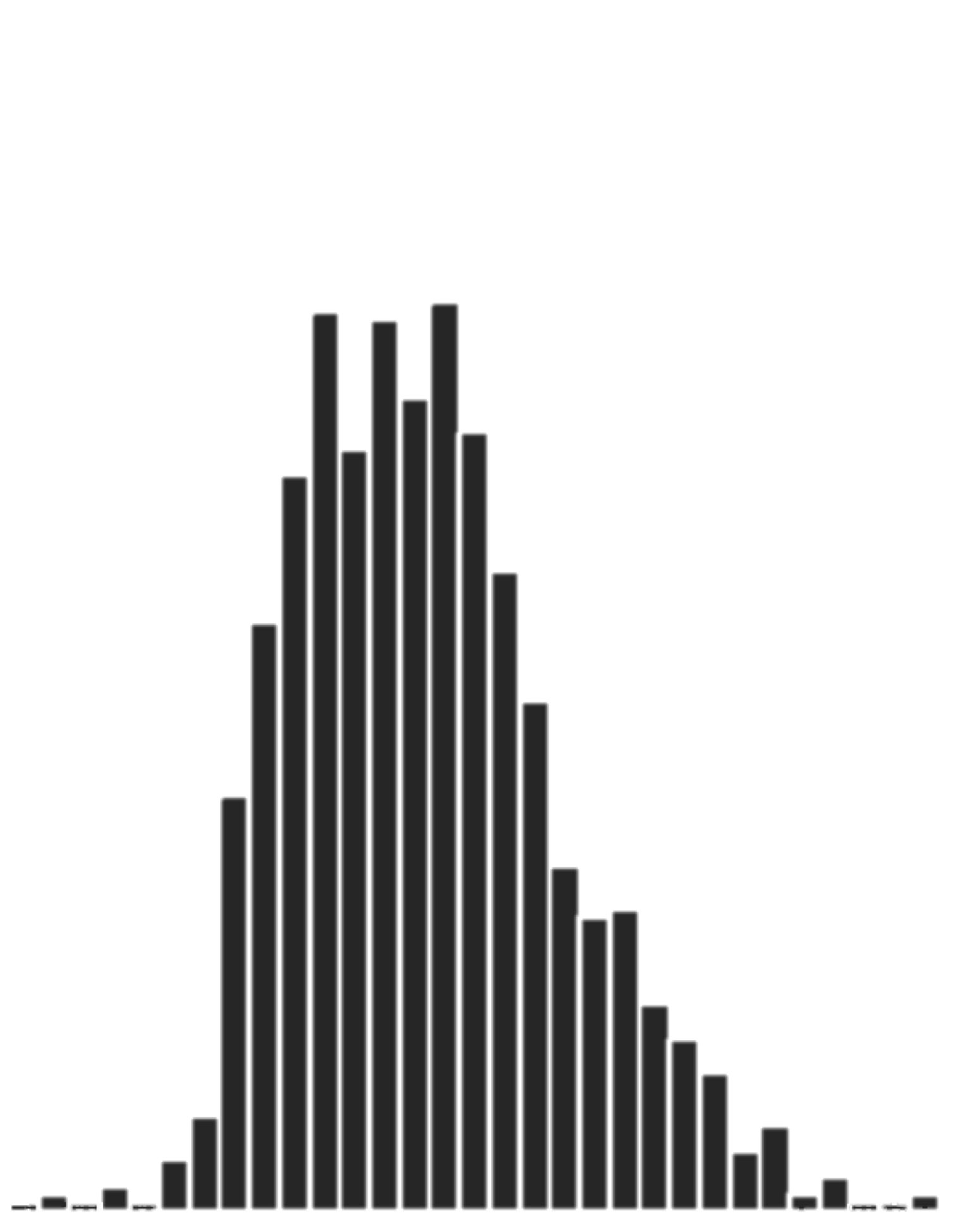}} & {\fontfamily{pag}\selectfont 0.57}\\
{\fontfamily{pag}\selectfont Negative Emotion} &  \raisebox{-.2\height}{\includegraphics[width=3.0cm,height=0.8cm]{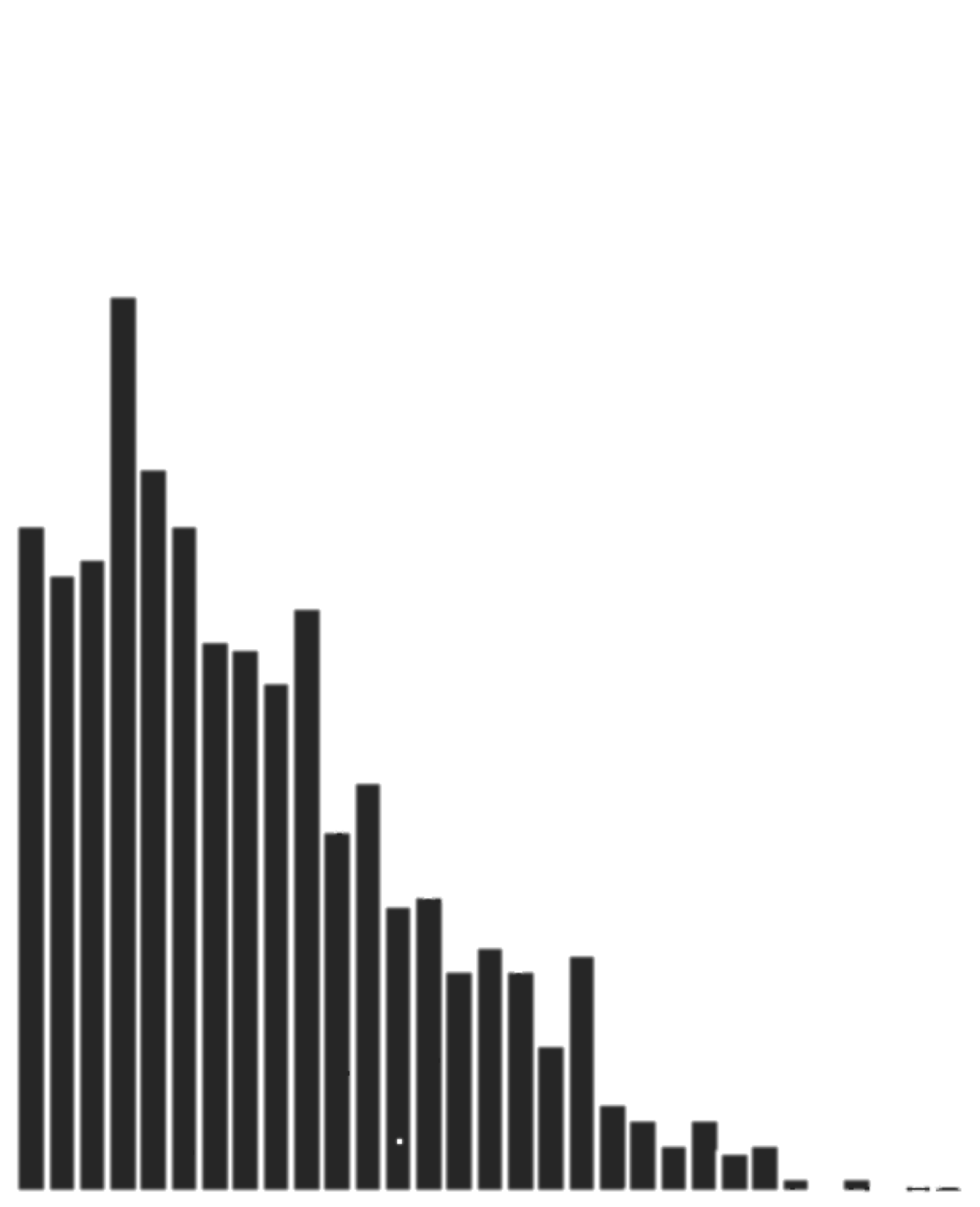}} & {\fontfamily{pag}\selectfont 0.53} & \raisebox{-.2\height}{\includegraphics[width=3.0cm,height=0.8cm]{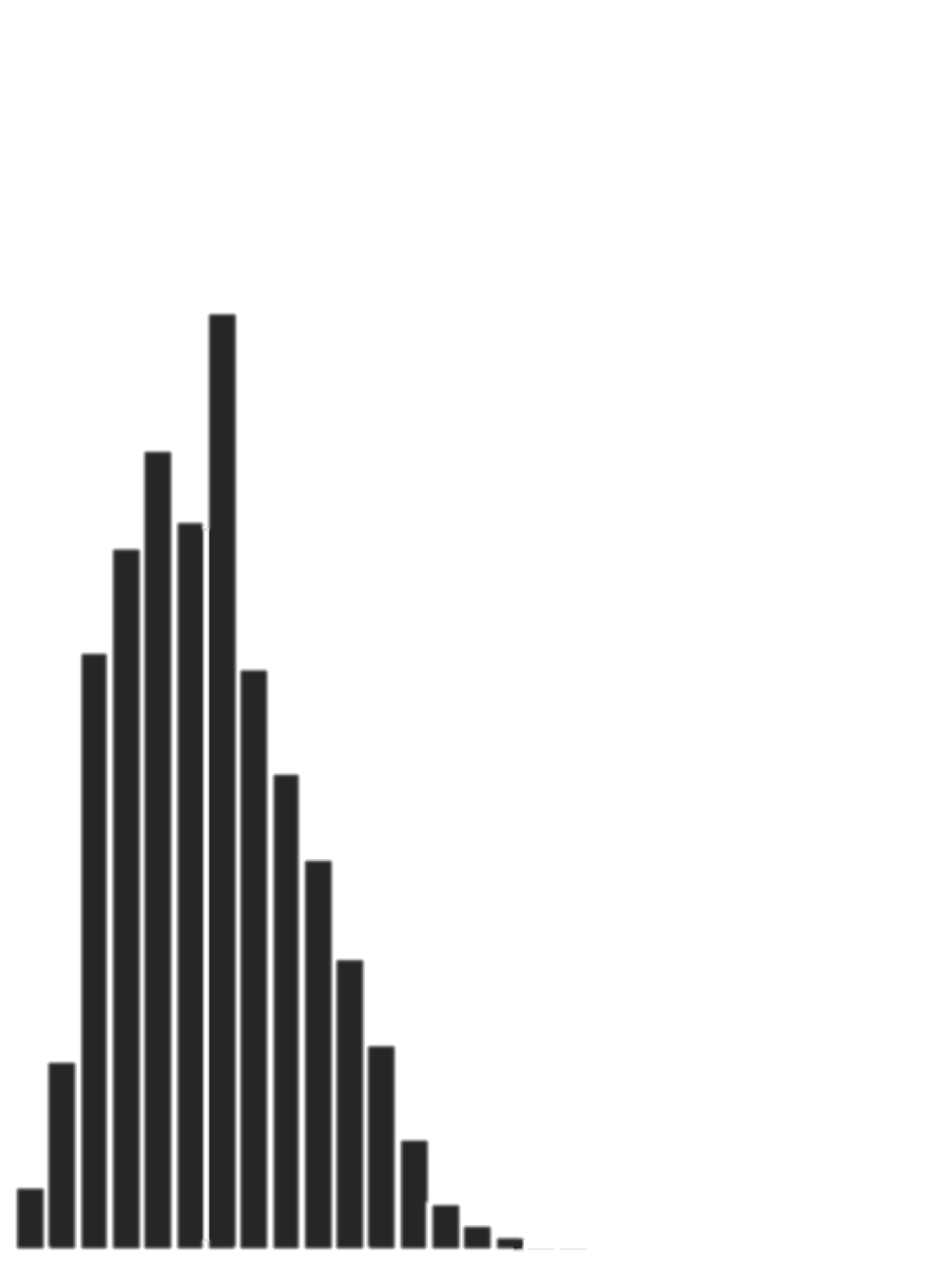}} & {\fontfamily{pag}\selectfont 0.46}\\
{\fontfamily{pag}\selectfont Likes per View} &  \raisebox{-.2\height}{\includegraphics[width=3.0cm,height=0.8cm]{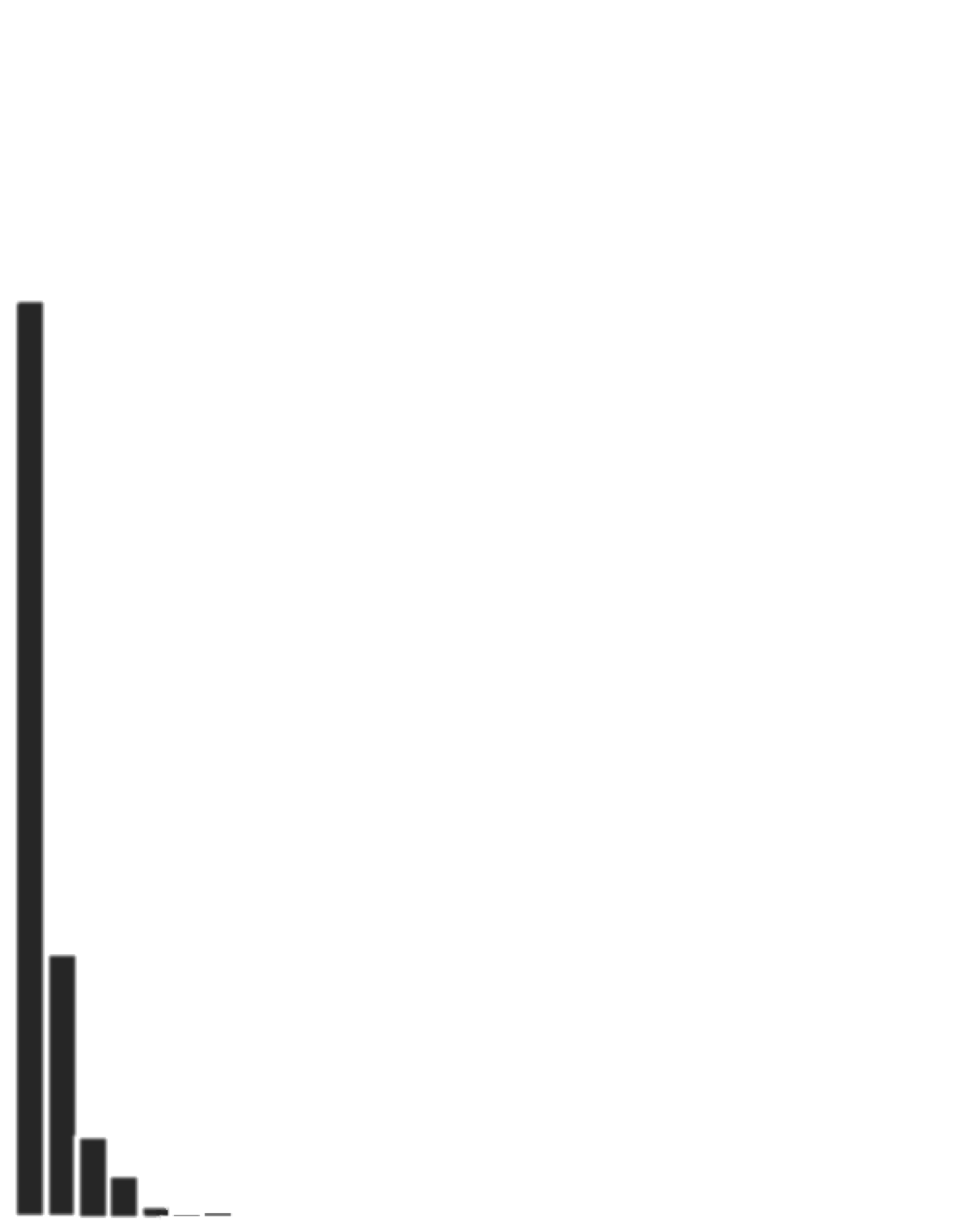}} & {\fontfamily{pag}\selectfont 0.98} & \raisebox{-.2\height}{\includegraphics[width=3.0cm,height=0.8cm]{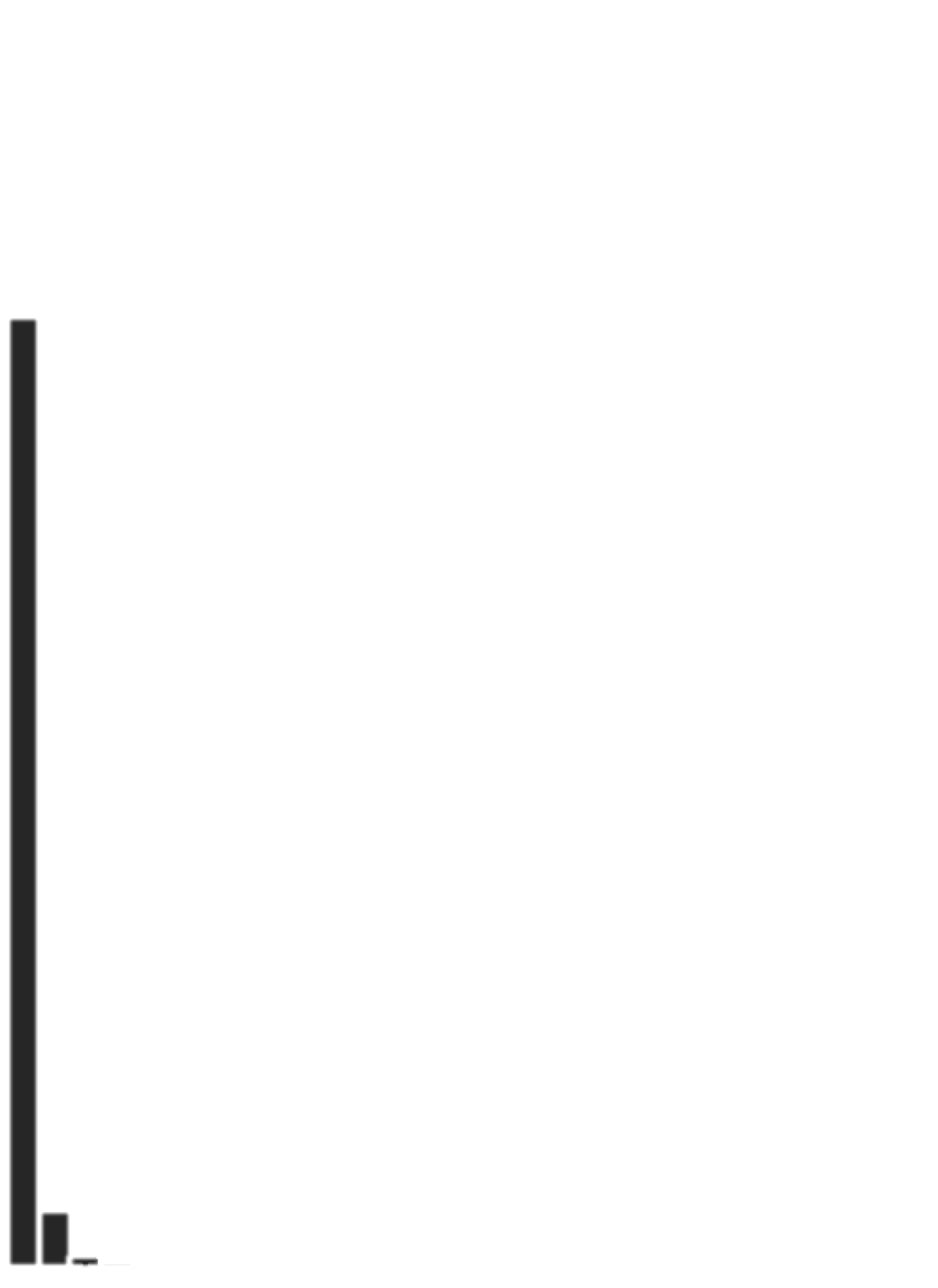}} & {\fontfamily{pag}\selectfont 0.86}\\
{\fontfamily{pag}\selectfont Dislikes per View} &  \raisebox{-.2\height}{\includegraphics[width=3.0cm,height=0.8cm]{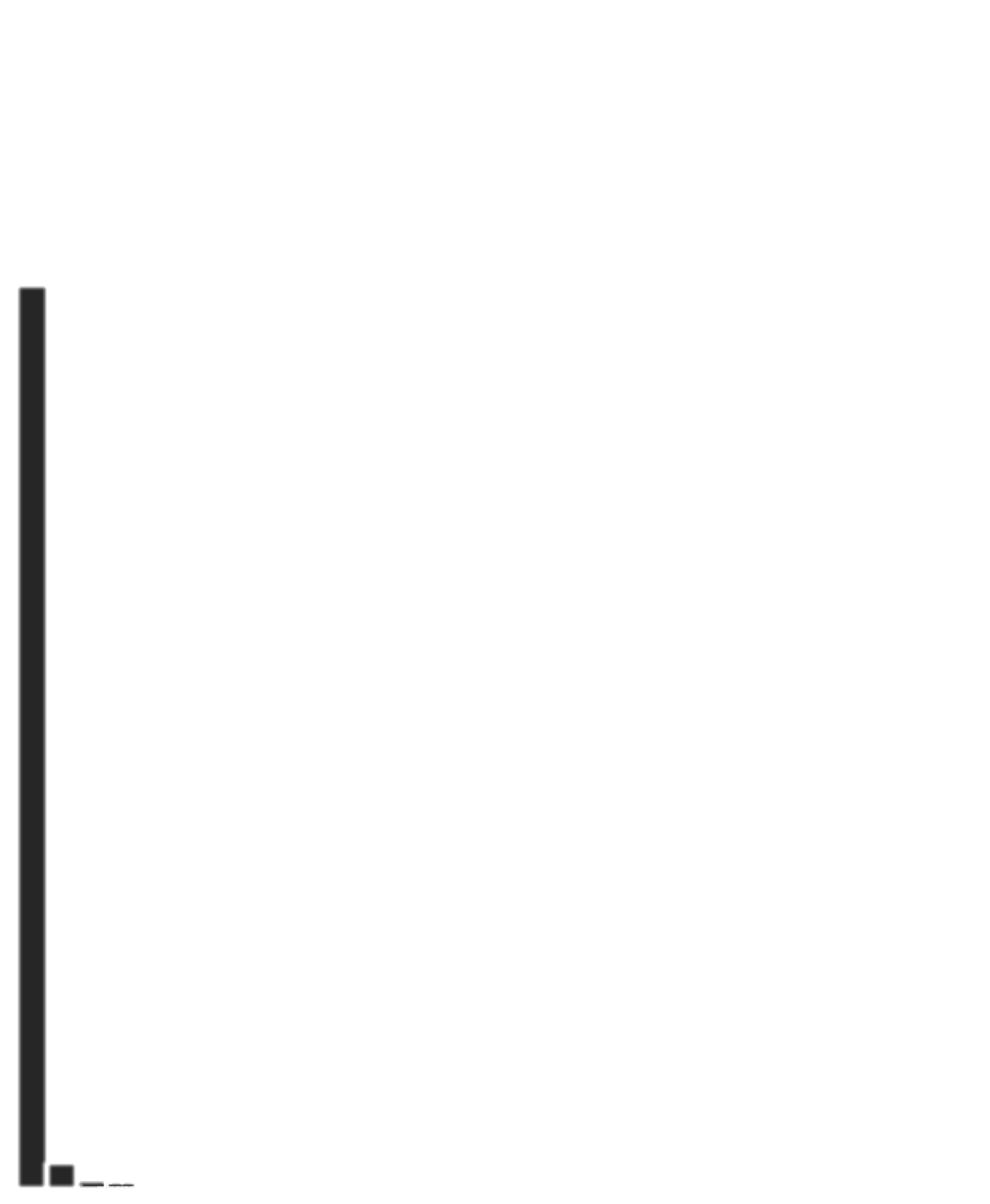}} & {\fontfamily{pag}\selectfont 0.26} & \raisebox{-.2\height}{\includegraphics[width=3.0cm,height=0.8cm]{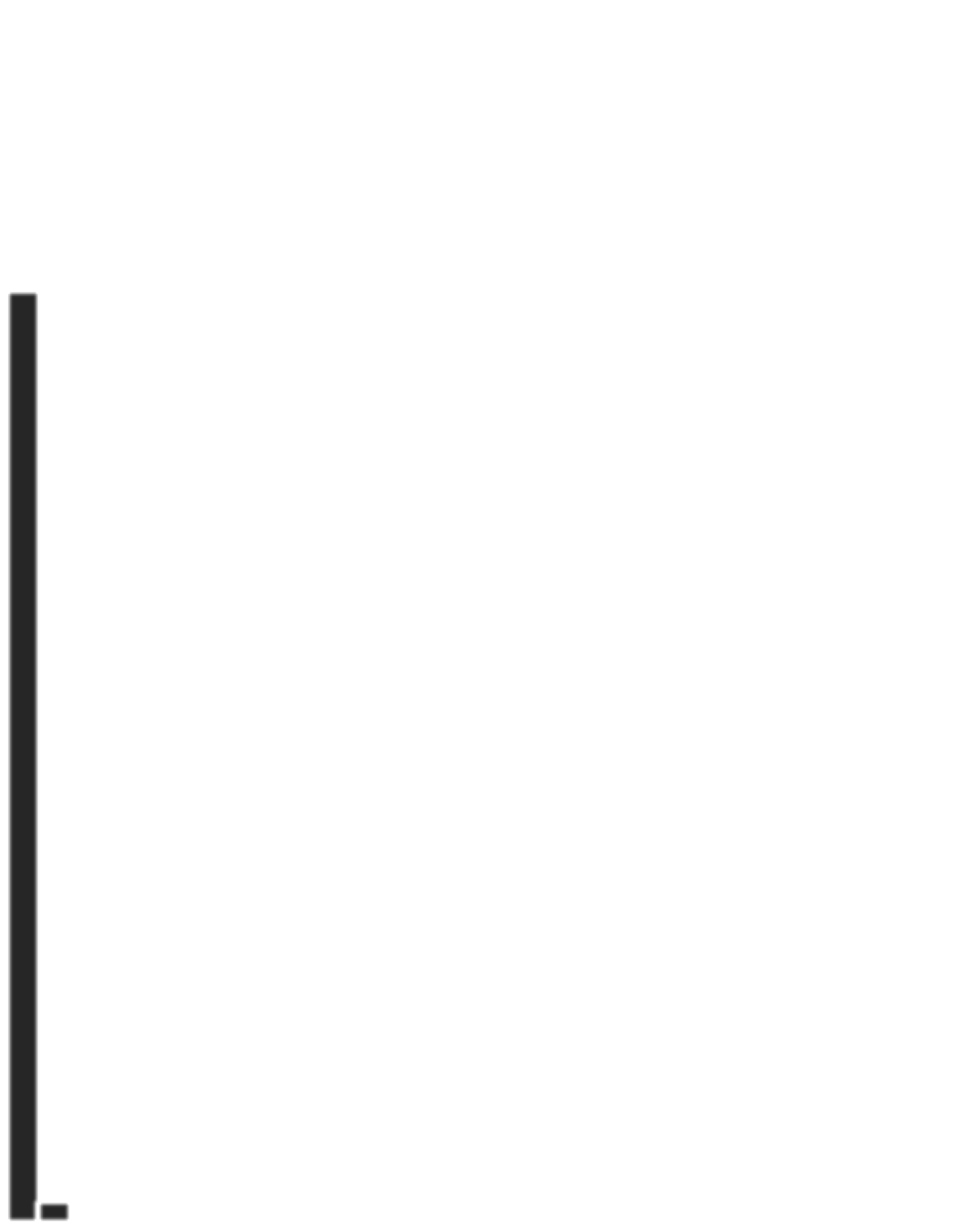}} & {\fontfamily{pag}\selectfont 0.18}\\
{\fontfamily{pag}\selectfont Like-Dislike Ratio} &  \raisebox{-.2\height}{\includegraphics[width=3.0cm,height=0.8cm]{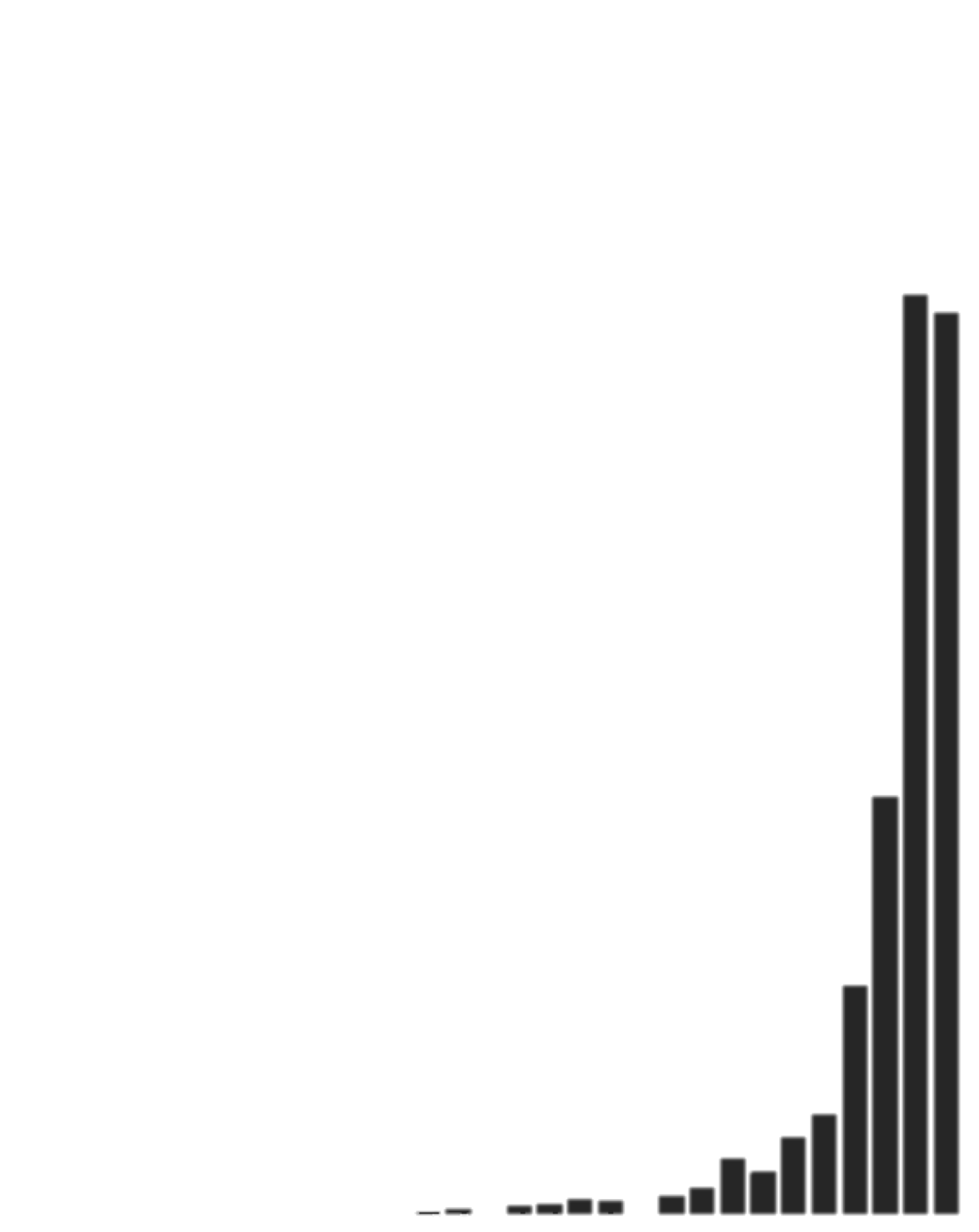}} & {\fontfamily{pag}\selectfont 1} & \raisebox{-.2\height}{\includegraphics[width=3.0cm,height=0.8cm]{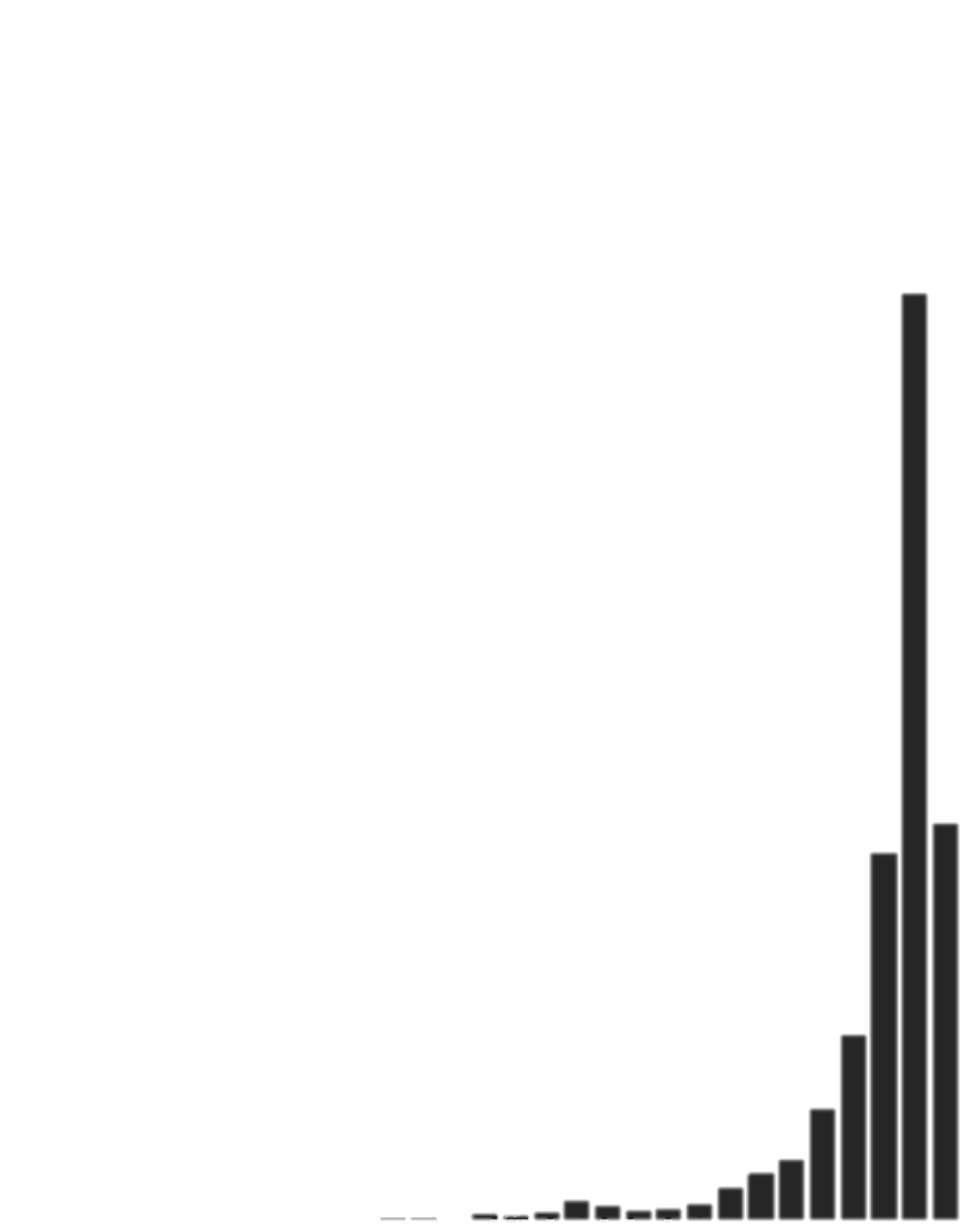}} & {\fontfamily{pag}\selectfont 1}\\
{\fontfamily{pag}\selectfont Shares per View} &  \raisebox{-.2\height}{\includegraphics[width=3.0cm,height=0.8cm]{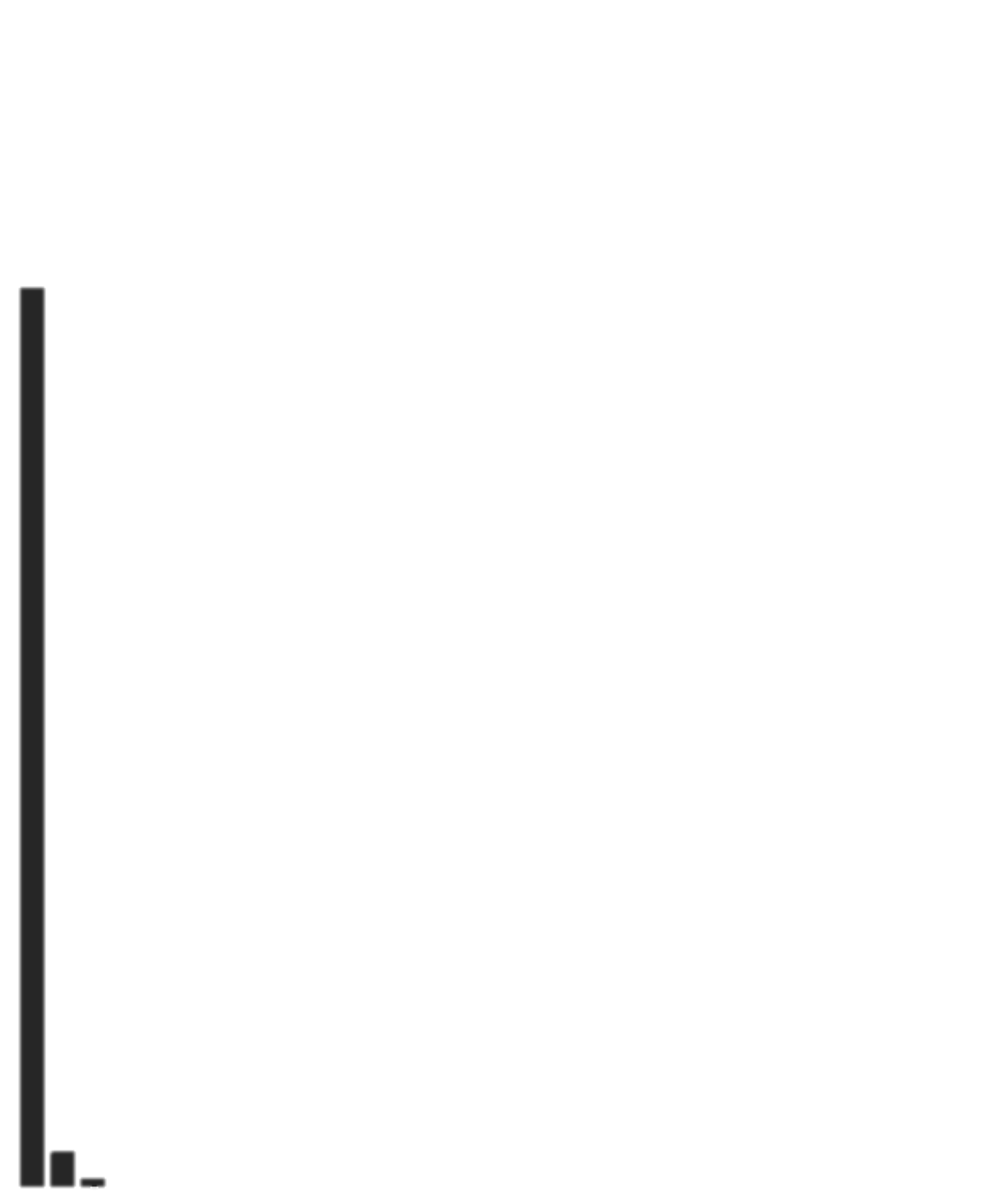}} & {\fontfamily{pag}\selectfont 0.18} & \raisebox{-.2\height}{\includegraphics[width=3.0cm,height=0.8cm]{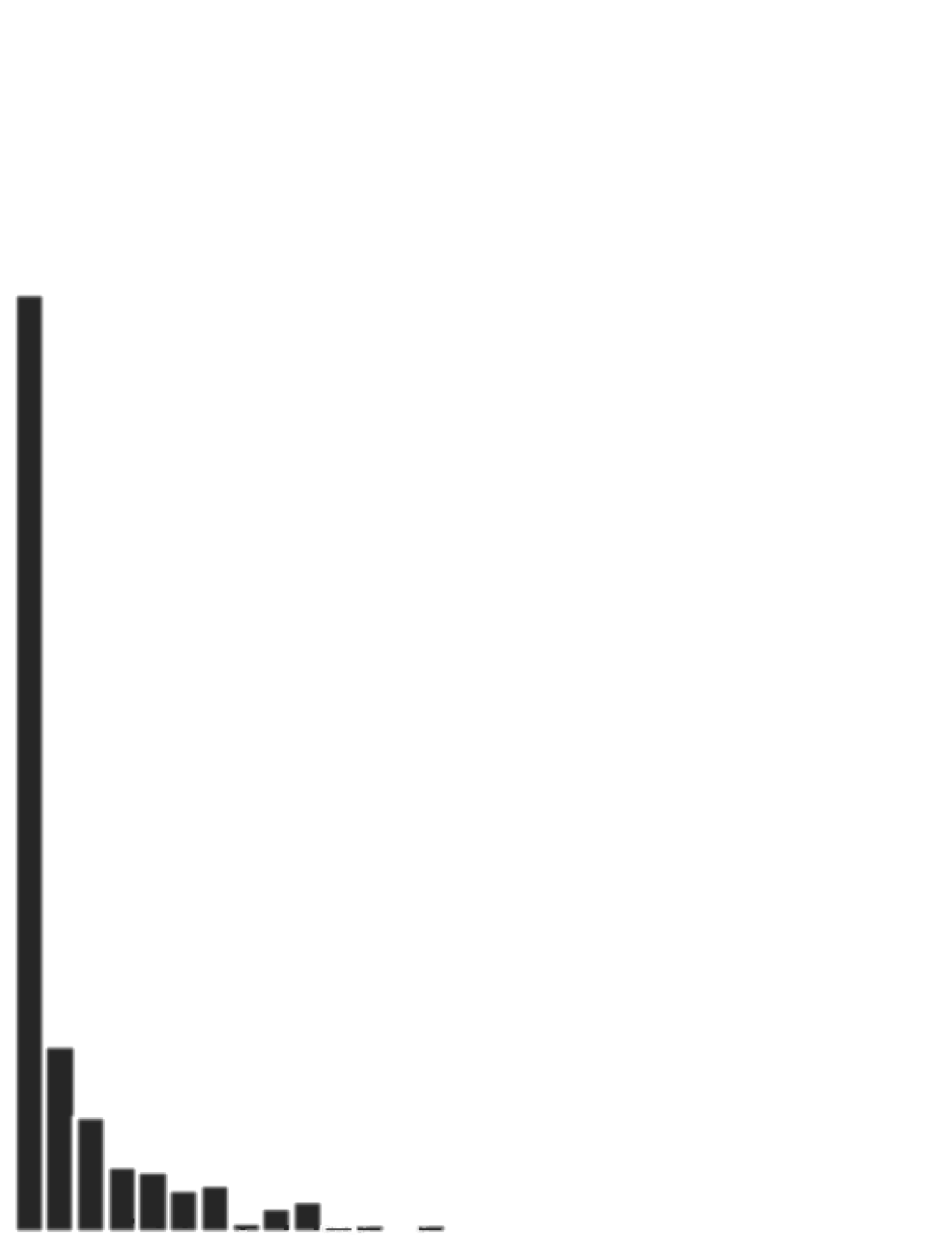}} & {\fontfamily{pag}\selectfont 0.016}\\
\\
\end{tabular}}

\caption{Distribution of user engagement and independent variables used to estimate the engagement in each dataset. The distributions accompanying each variable begin at zero and end at the adjacent maximum.}
\label{tab:hist}
\end{table}

\subsection{Sentiment Analysis and Mixed-effects Model}
In order to gauge the intensity and direction of emotionality of comments, we performed sentiment analysis  on all of the comments for a given video by using VADER, a lexicon and rule-based sentiment analyser~\cite{hutto2014vader}. This tool is specifically attuned to sentiment expressed in social media and sensitive both the polarity and the intensity. Thus, it is ideal for use with comments on YouTube which often use less formal language. We considered only positive and negative sentiment for our analysis.\footnote{We also computed emotionality of comments by Linguistic Inquiry and Word Count (LIWC)~\cite{tausczik2010psychological}. While LIWC is widely used, it does not include consideration for sentiment-bearing lexical items such as acronyms, initialisms, emoticons, or slang, which are known to be important for sentiment analysis of social media text. The correlations between sentiments computed by LIWC and Vader were 0.84 (positive) and 0.83 (negative) for the \emph{Individual Logs} dataset, and 0.82 (positive) and 0.58 (negative) for the \emph{Random Videos} dataset, all significant at $p<0.001$. We therefore focus on using VADER in this work.}

We performed two mixed-effects OLS regressions, one regression for each dataset. We slightly modified the dependent variable for the regression, using the proportion of view duration for each video (rather than the view duration itself). In other words, for the \emph{Random Videos} we used the average view duration for each video divided by the video length. For the \emph{Individual Logs} dataset we used the user dwell time on the video page divided by video length. 

In the models for both datasets, we added control variables capturing key video properties, video length and age (elapsed time since the video was published). Both these variables were added as fixed effects. The key video variables we examine in this work include view count, comments per view (the number of comments divided by the number of views), likes/dislikes/shares per view (similarly computed), and positive and negative sentiments in comments as computed by VADER and resulting in a 0--1 score. We also used discussion density (the proportion of comments that are replies to other comments) and like-dislike ratio since dense discussion and clear preference among users may help to explain view duration. We added interaction terms between the sentiment and other variables as well.

All independent variables except sentiments and like-dislike ratio were logged (base 10) to account for skewness before being standardized and centered. Multicollinearity between variables was not an issue as all variance inflation factors were less than~2. Table~\ref{tab:hist} presents the histogram distributions of variables used for our analysis. 

We used a mixed-effects model to control for multiple measurements of the same individuals (\emph{Individual Logs} dataset) and for differences between video in different categories (both datasets). Participants were added as a random effect in the individual logs model. We used the video category as a random effect as well in both models, because we noticed significant category-dependent differences in view duration. Indeed, in the \emph{Individual Logs} dataset, for example, we found that the categories Comedy, Entertainment, People \& Blogs, Film \& Animation, and Gaming, were positively associated with the watch duration dependent variable (these categories were watched significantly longer compared to the baseline of the Auto \& Vehicle category, even with video length and elapsed time as controls).

% ==================================================== %
%               Results                                %
% ==================================================== %

\section{Results}
\noindent Table~\ref{tab_regression} shows the results of the two regression models for the two different datasets. The intercepts for the regressions represent a point with all variables at zero, in which case view duration values are 49\% and 66\% of a video in the \emph{Random Videos} and \emph{Individual Logs} datasets, respectively (this disparity is indicative of the difference between the datasets). Note that in both models, the video length has a significant impact on the view duration: as expected, shorter videos were watched much longer in terms of proportion of the video viewed. Indeed, the video length accounts for much of the $R^2$ value (justifying our choice to use it as a control variable). However, we note that the $R^2$ of same models without the video length control was 0.19 for both. When excluding both control variables, the $R^2$ values were 0.13 (\emph{Random Videos}) and 0.18 (\emph{Individual Logs}). These results indicate that video popularity measures and sentiment of comments have significant explanatory power over user engagement.

\begin{table}[t]
\centering
\includegraphics[scale=0.66]{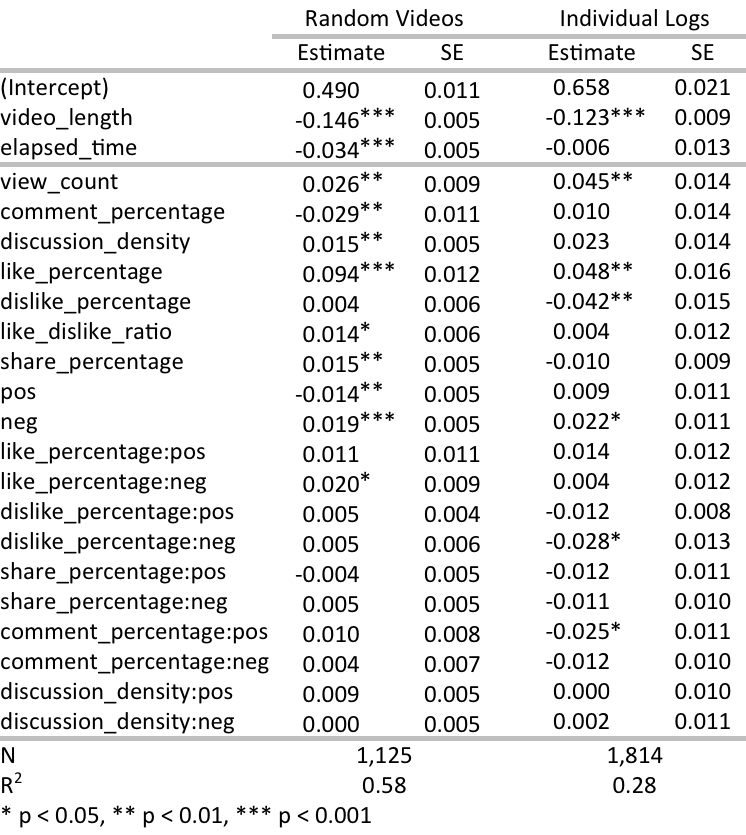}
\caption{Mixed-effects linear regressions estimating the percentages of average video watching (left; random videos) or video watching of individual users (right; individual logs).}
\label{tab_regression}
\end{table}

We treat the results that are highly significant across both datasets as our most robust findings. These results provide partial supports for both \emph{H1} and \emph{H2}. Supporting \emph{H1}, the video popularity and quality metrics of view count and likes per view were associated in both models with a higher portion of the video viewed (the ratio of people liking the video contributing more than the view count). Partially supporting \emph{H2}, the negative sentiment (but not the positive) in comments was positively correlated with view duration.
%in both models.

Other results were not significant in both models and we thus view them as less robust, perhaps indicative of potential association that can be exposed in a larger-scale study. The percentage of comments and positive sentiment (negative), and discussion density (positive) were correlated with view duration in the \emph{Random Videos} dataset. Dislike percentage was negatively correlated only in the \emph{Individual Logs} dataset. Finally, we note that there was no robust trend on Interaction terms.

% ==================================================== %
%       Discussion and Conclusion                      %
% ==================================================== %

\section{Discussion and Conclusion}
\noindent Drawing on two distinct YouTube datasets, this study demonstrates the relationship between popularity metrics, comment sentiment, and video watching (or the proportion of video watched by users). Our analysis shows a robust pattern: videos with higher ratio of likes per view, higher negative sentiment in comments, and higher view count are more likely to be watched longer. We note that the direction of the causal relationship between watch duration and other parameters, as well as the effect of the actual video quality, are not evident from the data. More generalizable engagement mechanisms may be exposed by a controlled experiment with a few selected videos. Theories such as excitation transfer~\cite{alhabash2015comment} can inform a stronger theoretical understanding of the relationship between these various video metrics. Yet, our data-driven approach helps to understand user engagement beyond view count and shows a potential to predict engagement level by using more readily available metrics. We intend to explore those of more detailed mechanisms in future work.

An interesting question is how the fact that the user can immediately observe other popularity metrics (e.g. the view count, available from the YouTube page) can \emph{change} the viewing behavior rather than just predict it. If users take view count or like/dislike ratio as aggregate opinion, these observable metrics may change their expectations of the video and hence their watching behaviors. Such relationships would be interesting to study in future research that explores whether and how other such factors, which we call `social traces' (e.g., previous views, but also demographic information of previous users or other measures of their engagement) contribute to different view behaviors. 

Finally, an interesting question is what precise mechanism links the negative sentiment in comments and view duration. For instance, video content that is more emotionally compelling may inspire high arousal negative emotions like anger and anxiety which are known to link to content sharing behaviors~\cite{berger2012makes}.

% ==================================================== %
%                Acknowledgements                      %
% ==================================================== %

\section{Acknowledgements}
{\small
\noindent We thank Oded Nov and the anonymous reviewers for their helpful feedback. This work is supported by a Google Social Interactions grant with additional support from AOL and NSF CAREER 1446374.}

% ==================================================== %
%                References                            %
% ==================================================== %

\bibliographystyle{aaai}
{\small
\bibliography{references}}
\end{document}